\documentclass{emulateapj}
\usepackage{lscape}

\shorttitle{AGN Spectral Decomposition}
\shortauthors{Vanden Berk et al.}

\begin{document}

\journalinfo{Accepted for publication in AJ}


\title{Spectral Decomposition of Broad-Line AGNs and Host Galaxies}


\author{
Daniel E.\ Vanden Berk\altaffilmark{1},
Jiajian Shen\altaffilmark{1},
Ching-Wa Yip\altaffilmark{2},
Donald P.\ Schneider\altaffilmark{1},
Andrew J.\ Connolly\altaffilmark{2},
Ross E. Burton\altaffilmark{2,3},
Sebastian Jester\altaffilmark{4},
Patrick B.\ Hall\altaffilmark{5}
Alex S.\ Szalay\altaffilmark{6},
John Brinkmann\altaffilmark{7}
}

\altaffiltext{1}{Department of Astronomy and Astrophysics, The Pennsylvania State University, 525 Davey Laboratory, University Park, PA 16802; danvb@astro.psu.edu.}
\altaffiltext{2}{Department of Physics and Astronomy, University of Pittsburgh, 3941 O'Hara Street, Pittsburgh, PA 15260.}
\altaffiltext{3}{Physics Department, Case Western Reserve University, 10900 Euclid Avenue, Cleveland, OH 44106-7079}
\altaffiltext{4}{Fermi National Accelerator Laboratory, P.O.\ Box 500, Batavia, IL 60510}
\altaffiltext{5}{Department of Physics and Astronomy, York University, 4700 Keele St., Toronto, ON M3J 1P3, Canada}
\altaffiltext{6}{Department of Physics and Astronomy, Johns Hopkins University, 3701 San Martin Drive, Baltimore, MD 21218.}
\altaffiltext{7}{Apache Point Observatory, 2001 Apache Point Road, P.O.\ Box 59, Sunspot, NM 88349-0059.}


\begin{abstract}
Using an eigenspectrum decomposition technique, we separate the host
galaxy from the broad line active galactic nucleus (AGN) in a set of
$4666$ spectra from the Sloan Digital Sky Survey (SDSS), from
redshifts near zero up to about $0.75$.  The decomposition technique
uses separate sets of galaxy and quasar eigenspectra to efficiently
and reliably separate the AGN and host spectroscopic components. The
technique accurately reproduces the host galaxy spectrum, its contributing
fraction, and its classification.  We show how the accuracy of the
decomposition depends upon $S/N$, host galaxy fraction, and the
galaxy class.  Based on the eigencoefficients, the sample of SDSS
broad-line AGN host galaxies spans a wide range of spectral types,
but the distribution differs significantly from inactive galaxies.
In particular, post-starburst activity appears to be much more common
among AGN host galaxies.  The luminosities of the hosts are much higher
than expected for normal early-type galaxies, and their colors become
increasingly bluer than early-type galaxies with increasing host luminosity.
Most of the AGNs with detected hosts are emitting at between $1\%$ and
$10\%$ of their estimated Eddington luminosities, but the sensitivity of the
technique usually does not extend to the Eddington limit.  There are
mild correlations among the AGN and host galaxy eigencoefficients,
possibly indicating a link between recent star formation and the onset
of AGN activity.  The catalog of spectral reconstruction parameters is
available as an electronic table.
\end{abstract}



\section{Introduction}

Quasars, and active galactic nuclei (AGNs) in general, are known to exist
at the centers of galaxies, and it is believed that accretion of galactic
material onto a supermassive black hole is the primary mechanism of their
energy generation \citep[e.g.][]{lyndenbell69, rees84}.  Numerous studies
now suggest a close link between the growth of supermassive black holes
and normal galaxies, in particular a correlation between black hole mass
and galactic stellar bulge mass \citep[e.g.][for a review]{ferrarese02}.
An understanding of the physical connection between the AGN central engine
and host galaxy requires a description of the properties of both the AGN
and host separately, and in relation to one another, across a wide range
of intrinsic properties.  However, it is observationally difficult to
study the properties of both an AGN and its host galaxy.  At high AGN
luminosities ($M_{B}\lesssim-23$, traditionally defined as the quasar
regime), the contrast between the AGN and host galaxy is so great that
the host cannot be easily discerned from the AGN.  At low AGN luminosities,
the AGN cannot be observed without significant contamination from the host
unless very high spatial resolution techniques are employed.

Much effort has been applied to the development of techniques
for the spatial image decomposition of AGNs and host galaxies
\citep[e.g.][]{kuhlbrodt04,mclure00,wadadekar99}.  The results from
image decomposition studies generally show that AGNs are usually
found in bulge-dominated galaxies at the bright end of the luminosity
function \citep[e.g.][]{smith86, dunlop03, falomo04, floyd04}; this is
especially true for radio loud AGNs and AGNs with higher luminosities
\citep[e.g.][]{hamilton02, dunlop03}.  Multi-band image studies
also show that the stellar populations of the host galaxies tend
to be bluer, and therefore are likely to be younger, than inactive
galaxies with the same morphological and luminosity characteristics
\citep{hutchings02, sanchez04, jahnke04a}.  There is also evidence
for mergers or other interactions in some of the host galaxy images
\citep{hutchings87,bahcall97,malkan98,marquez01,mcleod01,sanchez03,
sanchez04}, although it is not clear that host galaxies are more likely to
be interacting than their inactive counterparts \citep{schade00,dunlop03}.

Spectroscopic studies of AGN host galaxies have usually been confined to
the outer parts of the galaxies, well away from the contaminating nuclear
regions \citep{boroson82, balick83, boroson84, boroson85, hutchings90,
nolan01, miller03}.  Some integrated-field spectroscopy and other
spectroscopic AGN removal techniques have also been performed on a small
number of AGN host galaxies \citep{courbin02,jahnke04b}.  These studies
have revealed both old \citep{nolan01} and young \citep{hutchings90}
stellar populations.  Results supporting the interpretation of host
galaxies as massive bulge-dominated systems with relatively young stellar
populations were found by \citet{kauffmann03}, who analyzed the {\em
nuclear} spectroscopic properties of over $20,000$ narrow-line (type
2) AGNs found in the Sloan Digital Sky Survey \citep[SDSS][]{york00}.
They found that nearly all type 2 AGNs reside in massive galaxies
with ``early-type'' structural properties, but that the age, starburst
fraction, and ionization state of the AGN depends on the [O\,{\sc iii}]
emission luminosity.  Because type~2 AGN spectra contain little or no
continuum component, \citet{kauffmann03} were able to separate cleanly
the stellar from emission line contributions to the spectra using pure
stellar galaxy templates (they did not need to remove the broad-line
components of AGNs).

Recently, \citet{hao05a} and \citet{dong05} demonstrated that galaxy
stellar and emission line spectral components can be separated in
narrow-line and some broad-line AGNs by using galaxy eigenspectra
and a power-law continuum for a possible AGN continuum component.
The technique proved effective for isolating the emission line
components (from the galaxy, a possible AGN, or both) that were then
used to classify the galaxy as star forming or AGN \citep{hao05a}.
Spectral principal component analysis (PCA) has been performed on both
galaxy \citep{yip04a} and quasar \citep{yip04b} samples from the SDSS.
These studies showed that galaxies and quasars can be classified based
on only two or three eigencoefficients.  At low luminosities, the quasar
second eigenspectrum has a strong galactic component resembling the
first galaxy eigenspectrum.  These results suggest that a combination
of both galaxy and quasar eigenspectra may be effective in separating
the components of composite AGN/galaxy spectra.

In this paper we describe such a technique, and show that the galaxy
and AGN components can be reliably disentangled over a wide range of
galaxy types and galaxy to AGN flux fractions, even in spectra with modest
signal-to-noise ratios.  The technique has already been applied to a small
sample of SDSS AGNs by \citet{strateva05}, who used it to estimate AGN
continuum luminosities free of host galaxy components.  Here we apply the
technique to over $11,000$ AGN spectra from the SDSS third data release
\citep[DR3,][]{abazajian05}.  The properties of the AGNs and host galaxies
can be studied separately or in relation to each other.  In addition,
the large sample size makes it possible to examine population statistics
of broad-line AGNs and host galaxies in ways not possible previously.

The focus of this paper is on broad-line (``type 1'') objects,
which we will generically refer to as AGNs, regardless of luminosity.
Quasars are defined as broad-line AGNs with luminosities brighter than
$M_{i}=-22.0$ \citep[following the definition given in the SDSS quasar
catalog by][]{schneider05}.

The SDSS AGN survey and dataset are described in \S\,2.
The eigenspectrum spectral decomposition technique is described along with
the results of simulation tests in \S\,3.  The application
of the technique to the SDSS AGN spectra is described in \S\,4,
and some characteristics of the AGN and host populations are described
in \S\,5.  The results are discussed and summarized
in \S\,6.  Throughout the paper we assume a flat,
$\Lambda$-dominated cosmology with parameter values $\Omega_{m}=0.3,
\Omega_{\Lambda}=0.7$ and $H_{0}=70{\rm \,km\,s^{-1} Mpc^{-1}}$.

%
%
\section{The SDSS AGN Sample \label{S_dataset}}

\subsection{The Main SDSS Quasar Survey\label{S_mainSurvey}}

The broad-line AGNs used for this study were selected from the SDSS.
The SDSS is a project to image of order $10^{4}\,{\rm deg}^{2}$ of sky,
mainly in the northern Galactic cap, in five broad photometric bands
($u,g,r,i,z$, \citep{fukugita96}) to a depth of $r \sim 23$, and to
obtain spectra of $10^{6}$ galaxies and $10^{5}$ quasars selected from
the imaging survey.  Imaging observations are made with
a dedicated 2.5m telescope at Apache Point Observatory in New Mexico,
using a large mosaic CCD camera \citep{gunn98} in a drift-scanning mode.
Absolute astrometry for point sources is accurate to better than $100$
milliarcseconds rms per coordinate \citep{pier03}.  Site photometricity
and extinction monitoring are carried out simultaneously with a dedicated
20-inch telescope at the observing site \citep{hogg01}.  An assessment of
the image data quality is given by \citet{ivezic04}.

The imaging data are reduced and calibrated using the {\tt photo}
software pipeline \citep{lupton01}.  In this study we use three types
of magnitudes, two derived from imaging data, and one from spectroscopic
data.  The first magnitude we use is the point-spread function (PSF)
magnitude, which is derived from a Gaussian fit to the object and a
series of aperture corrections using a spatially varying PSF model
\citep{stoughton02}.  We also use ``cmodel'' magnitudes
\citep{abazajian04}, which are derived from a linear combination of
de~Vaucouleurs and exponential profile fits to object images.  The SDSS
photometric system is normalized so that the $u,g,r,i,z$ magnitudes
are approximately on the AB system (Oke \& Gunn 1983; Fukugita et al.\
1996; Smith et al.\ 2002; c.f.\ discussion by Abazajian et al.\ 2004).
The photometric zeropoint calibration is accurate to better than $2\%$
(root-mean-squared) in the $g$, $r$, and $i$ bands, and to better than
$3\%$ in the $u$ and $z$ bands, measured by comparing the photometry
of objects in scan overlap regions.  Spectroscopic targets are selected
by a series of algorithms \citep[see][]{stoughton02}, and are grouped by
three degree diameter areas or ``tiles'' \citep{blanton03}.  Two fiber-fed
double spectrographs can obtain 640 spectra for each tile; for the main
survey each tile contains 32 sky fibers, and roughly 500 galaxies, 50
quasars, and 50 stars.  The wavelength range of the SDSS spectra covers
approximately $3800${\AA} to $9200${\AA} at a spectroscopic resolution
of about 1800.  The AGN spectra used here were corrected for Galactic
extinction using the reddening map constructed by \citet{schlegel98}, and
the average Milky Way extinction curve described by \citet{fitzpatrick99}.
``Spectroscopic magnitudes'' are derived by convolving the calibrated
spectra with the SDSS $g$, $r$, and $i$ filter transmission curves that
include 1.3 airmasses of extinction.

Quasar candidates are selected from the SDSS color space and unresolved
matches to sources in the FIRST radio catalog \citep{becker95},
as described by \citet{richards02}.  Quasars are also often
identified because the objects were targeted for spectroscopy by
non-quasar selection algorithms, such as optical matches to ROSAT
sources \citep{stoughton02,anderson03}, various classes of stars
\citep{stoughton02}, so-called serendipity objects \citep{stoughton02},
and galaxies \citep{strauss02}.  The images of quasar candidates selected
as likely ``low-redshift'' ($z\lesssim 3$) objects, are allowed to be either
resolved or unresolved.  The SDSS {\tt photo} pipeline classifies object
images as resolved or unresolved based on the difference between PSF
and cmodel magnitudes \citep{abazajian04} \citep[see also discussions
by][]{stoughton02,scranton02,strauss02}.  It is important for this
study that resolved objects can be quasar candidates, because a large
fraction of AGN spectra for which a host component can be detected have
extended image profiles (see \S\,4.1).  The completeness
of the SDSS quasar selection algorithm is close to $95\%$ up to the $i$
band limiting survey magnitude of $19.1$ \citep{vandenberk05}.

\subsection{Selection of SDSS AGN Spectra \label{S_selection}}
The primary set of quasars was selected from the catalog compiled
by \citet{schneider05}.  The catalog contains $46,420$ quasars, and
comprises what is believed to be nearly all of the verified quasars in
the SDSS third data release \citep{abazajian05}.  Quasars are defined
in the catalog to be those extragalactic objects with absolute PSF
$i$ band magnitudes brighter than $M_{i}=-22$, and with at least one
emission line having a FWHM larger than $1000 {\rm \,km\,s^{-1}}$.  The
absolute magnitudes were calculated from the observed PSF magnitudes
without any attempt to subtract possible host galaxy contributions.
Near the faint luminosity limit of the catalog, it is apparent
from inspection that many of the quasar spectra contain host galaxy
components.  The fraction of quasar spectra in the catalog with
clear galaxy components decreases with increasing quasar luminosity
(a point that will be quantified in \S\,4.1).

The limit of $M_{i}=-22$ (and most other traditional limits) for the
definition of a quasar is fairly arbitrary, and broad-line AGNs are
clearly present at fainter magnitudes \citep[e.g.][]{ho97}.  It is
expected that the host galaxy fraction is larger for the fainter AGNs.
For this study we have extended the catalog of \citet{schneider05}
by simply removing the absolute magnitude criterion.  The AGNs were
selected from the initial search list described by \citet{schneider05},
by inspecting all of the extragalactic objects that were rejected from the
quasar catalog because they did not meet the absolute magnitude limit.
All of the AGNs were still required to pass the emission line width
criterion.  This search resulted in a sample of 3584 low-luminosity
broad-line AGNs.

We will refer to the combined sample of quasars and low-luminosity
broad-line AGNs as the SDSS AGN sample.  The sample contains over $50,000$
broad-line AGNs.  Host galaxy components will not be detectable in the
majority of the AGN spectra.  At redshifts beyond $0.752$, the wavelength
range of the galaxy eigenspectra covered by the SDSS spectra is too small
for a reliable spectroscopic reconstruction (\S\,3.5).
The initial data set therefore consists only of those AGNs with
$z<0.752$.  Spectra were also rejected for further analysis if more
than $50\%$ of their pixels inside either of two wavelength regions
were flagged as potentially bad by the $\tt spectro$ pipeline (see
\citet{stoughton02} for a list of flags).  The two wavelength regions
are $4160 < \lambda < 4210${\AA}, where integrated flux densities are
measured (\S\,3.2), and the wavelength region at which an
AGN spectrum and the galaxy and quasar eigenspectra overlap, which
depends upon the redshift of the AGN.  After the redshift and good
pixel criteria are applied, the initial data set consists of $11,647$
AGN spectra.


\section{Eigenspectrum Decomposition of AGN and Host Galaxy Spectra
  \label{S_technique}}
The technique we employ to separate the spectroscopic components of
composite AGN and host galaxy spectra uses separate sets of quasar
and galaxy eigenspectra in linear combination.  One of the goals of
eigenspectrum analysis (often also called principal component analysis,
PCA, or Karhunen-Lo{\`e}ve transformation, KL) is to reduce the complexity
of a dataset (compression) by constructing a (hopefully) small set of
orthogonal eigenspectra that account for the bulk of the variations
within a sample of objects.  \citet{yip04a} and \citet{yip04b} showed
that the vast majority of the spectra of galaxies and quasars in the
SDSS could be described by only a relatively small number ($\lesssim
10$) of eigenspectra.  In this section we demonstrate that spectra
with significant contributions from both an AGN and a host galaxy can
be effectively decomposed using existing sets of quasar and galaxy
eigenspectra.

\subsection{The Quasar and Galaxy Eigenspectra \label{S_eigdescribe}}
The eigenspectra employed here are those described and made available
by \citet{yip04a} for the SDSS galaxy sample, and by \citet{yip04b}
for the SDSS quasar sample.  The amount of information contained in
each eigenspectrum is given by the relative amplitude of its eigenvalue.
The galaxy sample variance is strongly concentrated in just the first few
eigenspectra, with over $98\%$ of the information contained in the
first three eigenspectra \citep{yip04a}.  The quasar information is also
concentrated, but not as strongly as in the galaxy case, with $\approx
92\%$ of the sample variance accounted for by the first ten eigenspectra.
\citet{yip04b} showed that the resulting quasar eigenspectra depend on
both redshift and luminosity, which causes a reduction in the information
content of the top global eigenspectra (formed from the entire sample).
Eigenspectra formed from subsets of quasars separated into bins
of redshift and luminosity have a much more concentrated information
content; meaning that fewer of these ``local'' eigenspectra are required
to describe a quasar within a restricted redshift and luminosity bin.
It was also shown that eigenspectra from a given bin can be used to
reliably reconstruct spectra in adjacent bins, allowing redshift or
luminosity trends to be tracked consistently \citep{yip04b}.

For our application, most of the AGN/galaxy spectra will be at redshifts
$z<0.5$, so we primarily use the set of quasar eigenspectra in
the ``ZBIN 1'' low redshift bin, spanning $0.08 \le z < 0.53$, defined
by \citet{yip04b}.  It is also important that there is little or no
host galaxy contamination in the quasar eigenspectra, so that a galaxy
component is reconstructed only by galaxy (and not quasar) eigenspectra.
Therefore we use eigenspectra constructed from quasars in the ``C1'' high
luminosity bin (in the low redshift range), defined by \citet{yip04b},
which is sufficiently luminous that there is little evidence for any
host galaxy component in the top eigenspectra; this bin spans an absolute
magnitude range of $-24 > M_{i} \ge -26$.

Galaxy or quasar spectra can be reconstructed as linear combinations
of eigenspectra
\begin{eqnarray}
   f^{R}_{\lambda} = \sum_{k=1}^{m}a_{k}e_{k}(\lambda),
  \label{Eq_recon}
\end{eqnarray}
where $f^{R}_{\lambda}$ is the reconstructed flux density as a function
of wavelength $\lambda$, and the $a_{k}$ are the eigencoefficients
of the corresponding wavelength dependent eigenspectra $e_{k}(\lambda)$.
Because the information content of the eigenspectra is concentrated
in the first several modes, a relatively small number of eigenspectra
can be used to closely approximate a given spectrum, while at the
same time suppressing much of the spectral noise, and interpolating over
small gaps in the spectra \citep{connolly95,connolly99,yip04a}.

Following \citet{connolly95}, we define two angles, called classification
angles, which are formed from the first three galaxy eigencoefficients.
The first $\phi$, is the mixing angle of the first two eigencoefficients,
$a_{1}$ and $a_{2}$, of a galaxy spectrum
\begin{eqnarray}
  \phi = \tan^{-1}(a_{2}/a_{1}).
  \label{Eq_phi}
\end{eqnarray}
The second angle $\theta$, is formed from the third eigencoefficient
$a_{3}$, of a galaxy spectrum
\begin{eqnarray}
  \theta = \cos^{-1}(a_{3}).
  \label{Eq_theta}
\end{eqnarray}
The eigencoefficients are normalized such that
\begin{eqnarray}
  \sum_{k=1}^{m}a_{k}^{2} = 1\,,
  \label{Eq_norm}
\end{eqnarray} 
where $m$ is the total number of eigenspectra used for a reconstruction.
It was shown by \citet{connolly95} that the spectral type of a galaxy
is tightly correlated with $\phi$, while post-starburst activity can be
discriminated with $\theta$ \citep{connolly95,castander01}.  The ability
of the spectral decomposition method to accurately recover the galaxy
classification angles from spectra with a wide range of characteristics
is one of the primary tests of the technique. \citet{yip04b} showed
that the first two eigencoefficients of quasar decompositions can be
used to classify quasars (although the classification angle formed
from the eigencoefficients was not explicitly defined).  We define
the classification angles for both host galaxies and quasars/AGNs in
the same way, according to equations\,\ref{Eq_phi}--\ref{Eq_norm}, and
denote them with the subscripts $H$ and $A$ respectively: $\phi_{H},
\theta_{H}, \phi_{A}, {\rm \,and\,} \theta_{A}$.

To test the dependence of the spectral decomposition method on
signal-to-noise ratio and host galaxy fraction, galaxy and quasar
templates with known eigencoefficients were modified in controlled
ways and the effect on the resulting eigencoefficients were examined.
The galaxy templates we used are the six high-$S/N$ templates (labeled
$a$ through $f$) constructed by \citet{yip04a} from objects in specific
regions of the projected classification plane.  These were constructed
to correspond to the spectra of each normal galaxy type in the atlas of
nearby galaxies compiled by \citet{kennicutt92}.  Each of the templates
was reconstructed using the first five galaxy eigenspectra to obtain the
eigencoefficients.  A quasar template was made for this study by selecting
a representative spectrum from the C1 quasar redshift-luminosity bin, and
reconstructing it using the first 10 quasar eigenspectra.  The spectrum
was selected to have a high $S/N$, and inspected to ensure that there
was no indication of a host galaxy contribution.

%
\begin{figure}
  \plotone{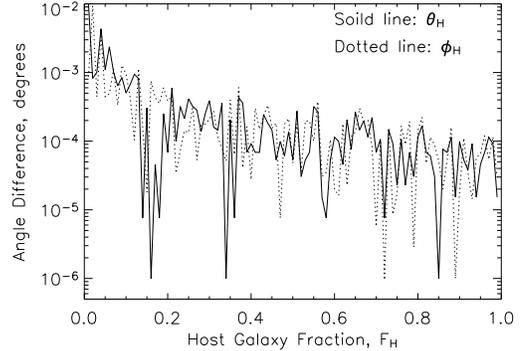}
  \caption{The difference between the true galaxy classification angles and
    best-fit angles as a function of the host galaxy fraction for noiseless
    input spectra.  The dotted line shows the $\phi_{H}$ differences,
    and the solid line shows the $\theta_{H}$ differences. The input
    spectra were constructed by combining the quasar template with the
    galaxy template from the $-12\degr < \phi_{H} < -8\degr$ range.
    \label{F_phithetaGNonoise}}
\end{figure}

\subsection{Recovery of Classification Angles from Noiseless Spectra
 \label{S_noiseless}}
One concern about the use of two separate sets of eigenspectra is that
they may not be orthogonal in combination.  That is, a composite 
host galaxy-quasar spectrum may not be uniquely described by a linear
combination of the eigenspectra from the two different bases.  While
\citet{yip04b} showed that the second global quasar eigenspectrum resembles
a galaxy spectrum, it still contains quasar features (e.g.\ some broad
emission line components), and the full galaxy component may be spread
across many eigenspectra.  Therefore, there is no guarantee that the
quasar and galaxy components of composite spectra can be reliably
reconstructed using the eigenspectrum method.

The reliability of the spectral reconstruction of composite host-quasar
spectra was first tested for the ideal case of noiseless spectra.
Galaxy template $d$ (from the range $-12\degr < \phi_{H} < -8\degr$ and
$80\degr < \theta_{H} < 100\degr$) was selected for this test because it
is derived from the region with the highest density of sample galaxies.
The reconstructed galaxy template was combined with the quasar template at
various levels of fractional contribution.  The fractional contribution
of the host galaxy to the composite spectrum $F_{H}$, is defined by
the integrated flux densities of the reconstructed quasar and galaxy
components over the rest wavelength range $4160 < \lambda < 4210${\AA},
\begin{eqnarray}
  F_{H} = \frac{\int_{4160}^{4210}
        {f^{R}_{\lambda,H}}\,d\lambda
        }{\int_{4160}^{4210}({f^{R}_{\lambda,A}}
        + {f^{R}_{\lambda,H}})\,d\lambda}.
  \label{Eq_FH}
\end{eqnarray}
The wavelength range was chosen because it avoids major galaxy stellar
absorption lines as well as strong quasar emission lines, and it is
covered by all of the SDSS spectra that can be usefully reconstructed.
The noiseless spectra were fit with the combined sets of quasar and galaxy
eigenspectra, using the first five galaxy and ten quasar eigenspectra,
and the classification angles $\phi_{H}, \theta_{H}, \phi_{A}$ and
$\theta_{A}$ were calculated.  (It is shown in \S\,3.3
that three galaxy and five quasar eigenspectra are sufficient for the
reconstruction, and using more does not change the results significantly.)
Figure \ref{F_phithetaGNonoise} shows the difference between the true
angles and best-fit angles as a function of the host galaxy fraction for
the noiseless input spectra.  It is evident that there is no significant
difference between the true and measured classification angles at any
level of host galaxy contribution.  Similar results were found for the
quasar classification angles.  The fitting technique reliably reconstructs
the galaxy and quasar components in the ideal case of noiseless spectra.

%
\begin{figure}
  \plotone{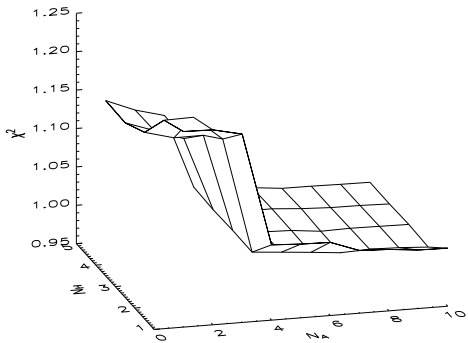}
  \caption{Average reduced $\chi^{2}$ values for reconstructed spectra
    compared to template spectra for varying numbers of quasar and
    galaxy eigenspectra.  The value of $\chi^{2}$ ranges from $1.2$
    at $N_{A}=1$ and $N_{H}=1$, to $1.0$ at $N_{A}=10$ and $N_{H}=5$.
    The average reduced $\chi^{2}$ value for each point on the grid was
    calculated from fits to 1000 simulated spectra. \label{F_NaNhChi}}
\end{figure}

\subsection{The Number of Eigenspectra \label{S_eignumber}}
The number of quasar and galaxy eigenspectra to use in the fitting
process is an important issue.  With too few eigenspectra, it will not
be possible to reconstruct all of the significant details in a spectrum.
The use of too many eigenspectra can lead to ``overfitting'' a spectrum,
so that eigenspectra of higher orders begin to fit noise features.

The optimal number of eigenspectra will clearly depend upon the quality
of a spectrum.  In \S\,3.4 we discuss the dependence of the
reconstructions on $S/N$ and host galaxy fraction.  Here we examine the
number of eigenspectra that should be used by fixing the $S/N$ and host
fraction at values typical of the spectra in the SDSS.  The same galaxy
and quasar templates used in the previous section were combined at a
galaxy fraction of $30\%$, and Gaussian noise was added to achieve a $S/N$
per pixel of $10$ (for SDSS spectra, this translates to a $S/N$ per {\em
resolution element} of $\approx 12$).  The template was fit using varying
numbers of both host galaxy and quasar eigenspectra, $N_{H}$ and $N_{A}$
respectively, and the quality of each fit was measured by calculating the
reduced $\chi^{2}$ value of the fit.  Figure\,\ref{F_NaNhChi} shows the
reduced $\chi^{2}$ values of the fits for each combination of the numbers
of quasar and galaxy eigenspectra.  The value of $\chi^{2}$ ranges from
$1.2$ at $N_{H}=1$ and $N_{A}=1$, to 1.0 at $N_{H}=5$ and $N_{A}=10$.
The $\chi^{2}$ value drops significantly with increasing numbers of
eigenspectra, until about three galaxy and five quasar eigenspectra,
at which point the values do not change greatly with the addition of
more eigenspectra.  Therefore, for the template used in this test,
there is no justification for using more than three galaxy and
five quasar eigenspectra.  For the AGNs in our sample, we have generally
used five galaxy and ten quasar eigenspectra, which are more than needed
in most cases, but not too many that the spectra are overfitted (the
reduced $\chi^{2}$ values do not drop significantly below unity).

The result that only a relatively small number of eigenspectra are
required is not unexpected.  The first three galaxy eigenspectra account
for over $98\%$ of the variation in SDSS galaxy spectra \citep{yip04a},
and the first 5 quasar eigenspectra in the low-redshift, high luminosity
bin account for $99.8\%$ of the SDSS quasar spectral variation in
that bin.  More subtle variations accounted for by higher orders are
often not apparent in spectra unless the $S/N$ level is very high.

%
\begin{figure}
  \plotone{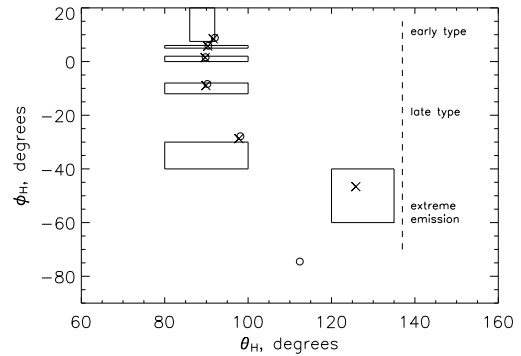}
  \caption{The measured galaxy classification angles for the six noiseless
  templates, using $N_{H}=5$ eigenspectra (circles), and $N_{H}=10$
  eigenspectra ($\times$'s).  The boxes show the regions from which
  galaxy spectra were selected to form the templates.  Labels show the
  general trend of galaxy type with $\phi_{H}$. \label{F_phithetaTemp}}
\end{figure}

While a small number of eigenspectra is sufficient for the most
common galaxy types, a larger number will be necessary for rare
galaxy types, particularly those with very strong emission lines
\citep{yip04a}. To illustrate this, Fig.\,\ref{F_phithetaTemp} shows
the measured classification angles for all six galaxy templates using
both $N_{H}=5$ and $N_{H}=10$ galaxy eigenspectra.  No noise or quasar
component was added in this test.  The measured classification angles
are almost indistinguishable in the two cases for all five templates with
$\phi_{H} > -40\degr$.  However there is a difference of about $30\degr$
in $\phi_{H}$ and about $15\degr$ in $\theta_{H}$ between the measurements
for the extreme emission line galaxy template.  The measurement made
using the larger number of eigenspectra is clearly better in that case.
Because of this, in cases with an initial $N_{H}=5$ measurement of
$\phi_{H} < -40\degr$, the spectra used in this study were 
decomposed a second time using $N_{H}=10$ host eigenspectra.

\subsection{Dependence on Signal-to-Noise and Host Galaxy
  Fraction\label{S_snFrac}}

The reliability of the component reconstruction clearly depends upon
the $S/N$ of a spectrum and the relative contribution of each component.
To quantify this dependence, we tested the routine on simulated spectra
generated as described above but with varying amounts of noise and host
galaxy contribution.  For these tests, the quasar template was combined
with the $-12\degr < \phi_{H} < -8\degr$ galaxy template.  The galaxy template
fraction $F_{H}$ was varied from near zero to near one.  The spectroscopic
noise level was varied to achieve a $S/N$ per pixel range between 5 and 30.
Random noise was added to each composite template spectrum according to
the desired $S/N$ level. The spectra were decomposed using $N_{H}=5$ and
$N_{A}=10$ eigenspectra, and the classification angles were measured.
This was repeated 1000 times for each composite template spectrum, and
the rms dispersion of the $\phi$ and $\theta$ values were calculated
for each combination of galaxy fraction and $S/N$.

%
\begin{figure*}
  \plottwo{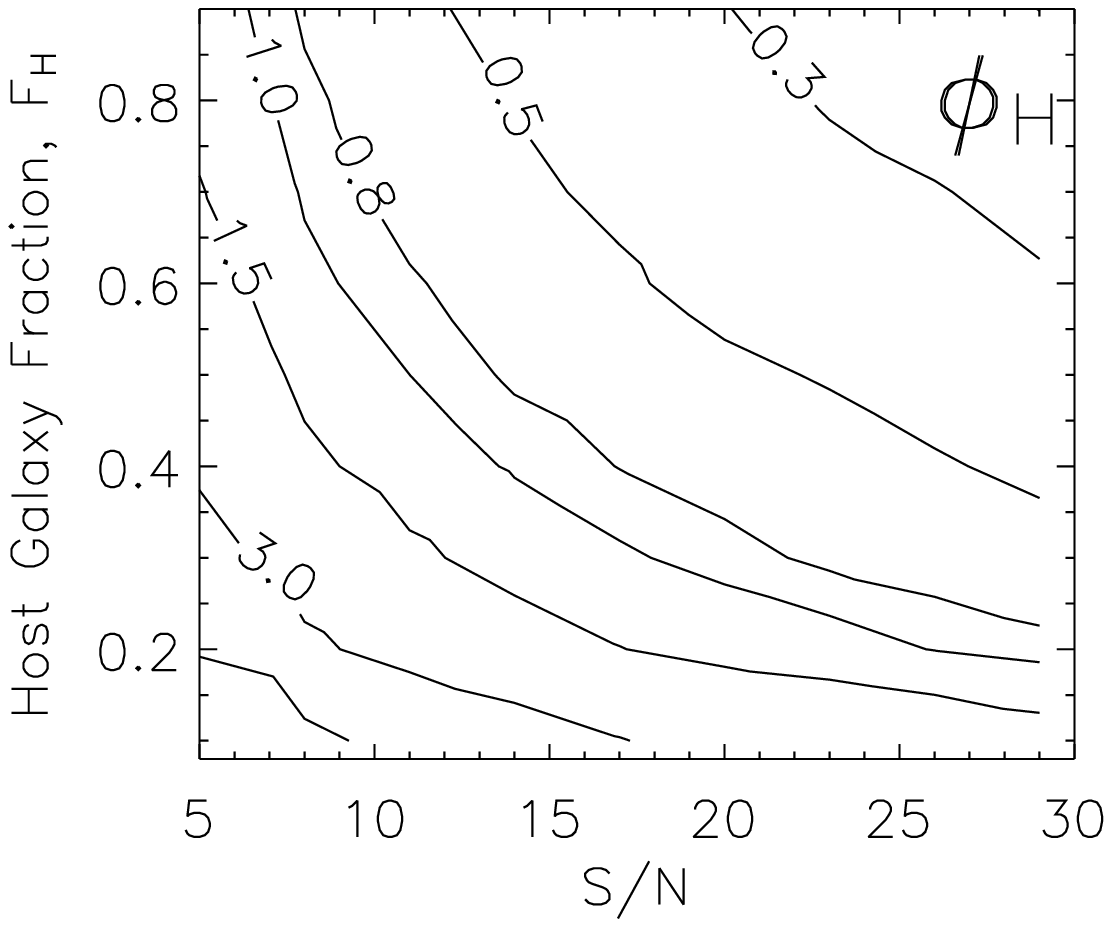}{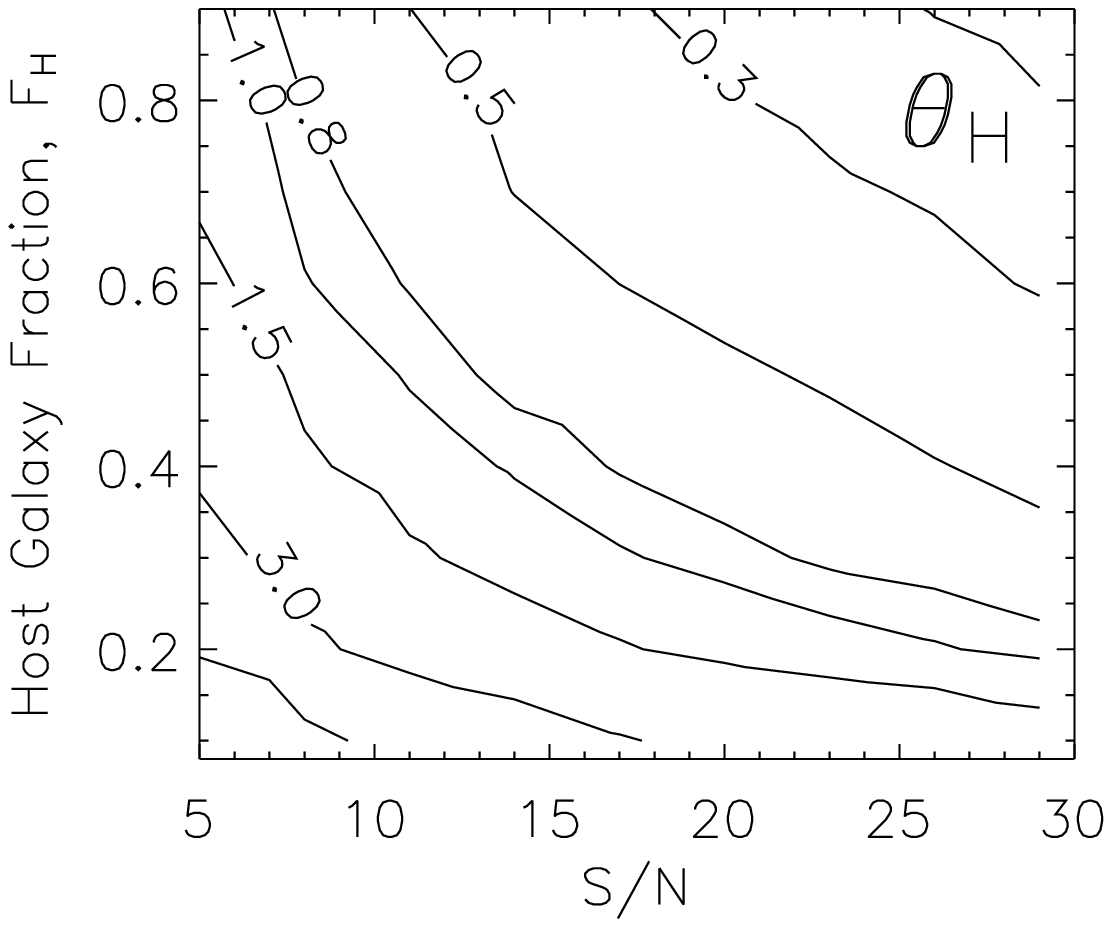}
  \caption{The rms dispersion, in degrees, on the measurement of $\phi_{H}$
    (left) and $\theta_{H}$ (right)
    for simulated spectra with varying levels of $S/N$ and host galaxy
    contribution fraction.  The contours are labeled by the value of the
    dispersion. \label{F_phiThetaGDisp}}
\end{figure*}

%
\begin{figure}
  \plotone{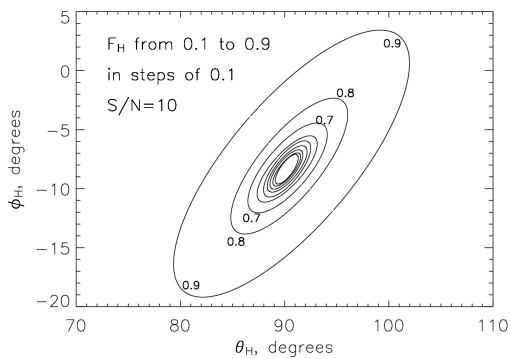}
  \caption{Confidence contours encircling $90\%$ of the results of
    reconstructions to the simulated spectra, for varying values of
    $F_{H}$.  The host galaxy fraction in the simulated spectra was varied
    from 0.1 to 0.9 in steps of 0.1, and the $S/N$ level was fixed to be
    10. The outer three contours are labeled by the values of $F_{H}$
    in the simulations.  The angle $\phi_{H}$ is generally correlated
    with star formation rate, while $\theta_{H}$ appears to be related
    to post-starburst activity. \label{F_phithetaEllipses}}
\end{figure}

%
\begin{figure}
  \plotone{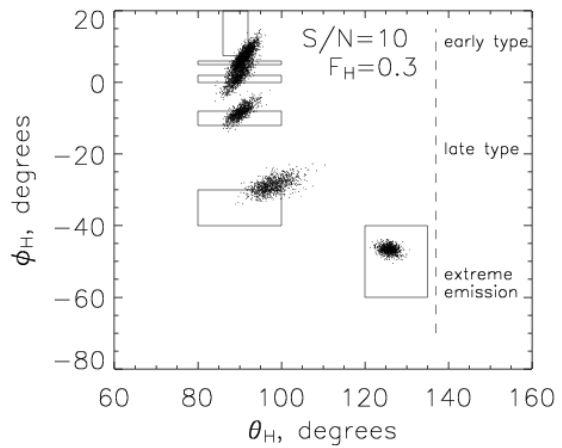}
  \caption{The classification angles for the six host galaxy templates with
    1000 simulations of added noise.  The $S/N$ level of the simulated
    spectra was fixed at 10, and the host galaxy fraction was fixed at
    $F_{H}=0.3$.  The boxes show the regions from which \citet{yip04a}
    selected galaxy spectra to construct the templates.  Labels show the
    general trend of galaxy type with $\phi_{H}$. \label{F_phitheta6temps}}
\end{figure}

The results for the dispersion of the $\phi_{H}$ and $\theta_{H}$
measurements are shown in Fig.\,\ref{F_phiThetaGDisp}.  As expected,
the dispersion decreases for increasing $S/N$ and galaxy fraction.
More interesting is the fact that even at modest $S/N$ and galaxy fraction
levels (say $S/N=10$ and $F_{H}=0.3$), the dispersions are only a few
degrees for each angle.  That is small enough to distinguish almost all
of the galaxy templates from each other.  Figure~\ref{F_phithetaEllipses}
shows the $90\%$ confidence ellipses (containing $90\%$ of 1000 random
realizations) in the $\phi_{H}-\theta_{H}$ plane for a range of values of
$F_{H}$, for simulations using the $-12\degr < \phi_{H} < -8\degr$ galaxy
template, with $S/N=10$.  The values of $\phi_{H}$ and $\theta_{H}$
are clearly covariant, but the confidence ellipses shrink rapidly
with increasing host galaxy fraction.  There are also no systematic
offsets in the values of the angles at any galaxy fraction.  Finally,
Fig.\,\ref{F_phitheta6temps} shows the measured values of $\phi_{H}$
and $\theta_{H}$ for the 1000 simulations of all six templates with
$S/N=10$ and $F_{H}=0.3$.  (The points are not centered in the middles
of the boxes because the distributions of real galaxies are not uniform
across the boxes; see \citet{yip04b}.)  All of the galaxy templates
can be distinguished from each other, except for the two with $\phi_{H}
> 5\degr$.

%
\begin{figure*}
  \plottwo{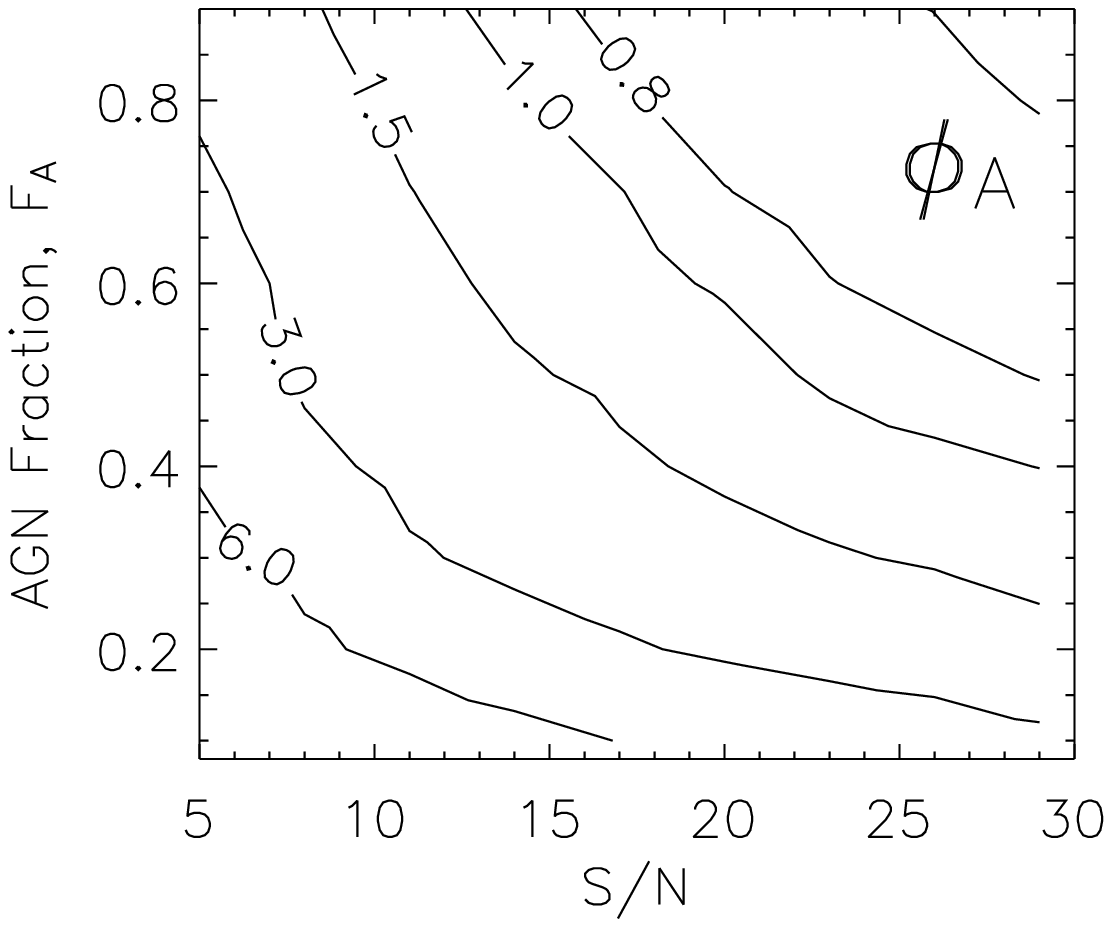}{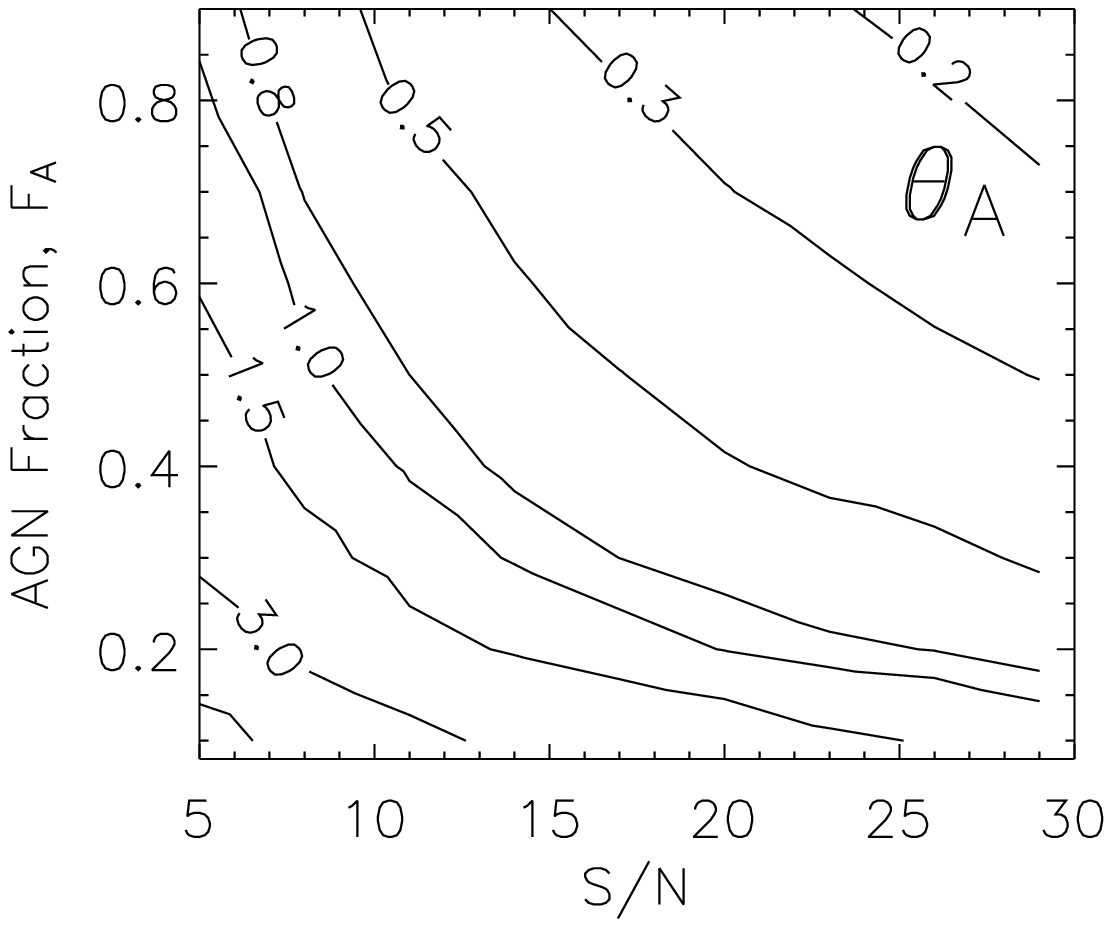}
  \caption{The rms dispersion, in degrees, on the measurement of $\phi_{A}$
    (left) and $\theta_{A}$ (right) for simulated spectra with varying
    levels of $S/N$ and host galaxy contribution fraction.  The contours
    are labeled by the value of the dispersion.  \label{F_phiThetaQDisp}}
\end{figure*}

Similar results were found for the quasar reconstructions.
Figure\,\ref{F_phiThetaQDisp} shows the dispersions of the $\phi_{A}$ and
$\theta_{A}$ measurements, as functions of both $S/N$ and quasar fraction.
The $\phi_{A}$ and $\theta_{A}$ dispersions are roughly twice as large
as the corresponding galaxy values.

\subsection{Dependence on Eigenspectrum Wavelength Coverage \label{S_wlcover}}

The eigenspectra cover the wavelength ranges spanned by the spectra
used to construct them.  For the eigenspectra used here, the rest frame
wavelength range is effectively $3500 - 7000${\AA}.  However, as   
redshift increases, a decreasing fraction of the eigenspectrum
wavelength range can be used, because rest frame wavelengths are
shifted out of the observed range.  The redshift at which there
is no more usable eigenspectrum wavelength coverage is clearly an
upper limit on the technique; however, a practical limit is reached
at lower redshifts as the reliability of the reconstructed spectra
decreases with shorter wavelength coverage.

%
\begin{figure}
  \plotone{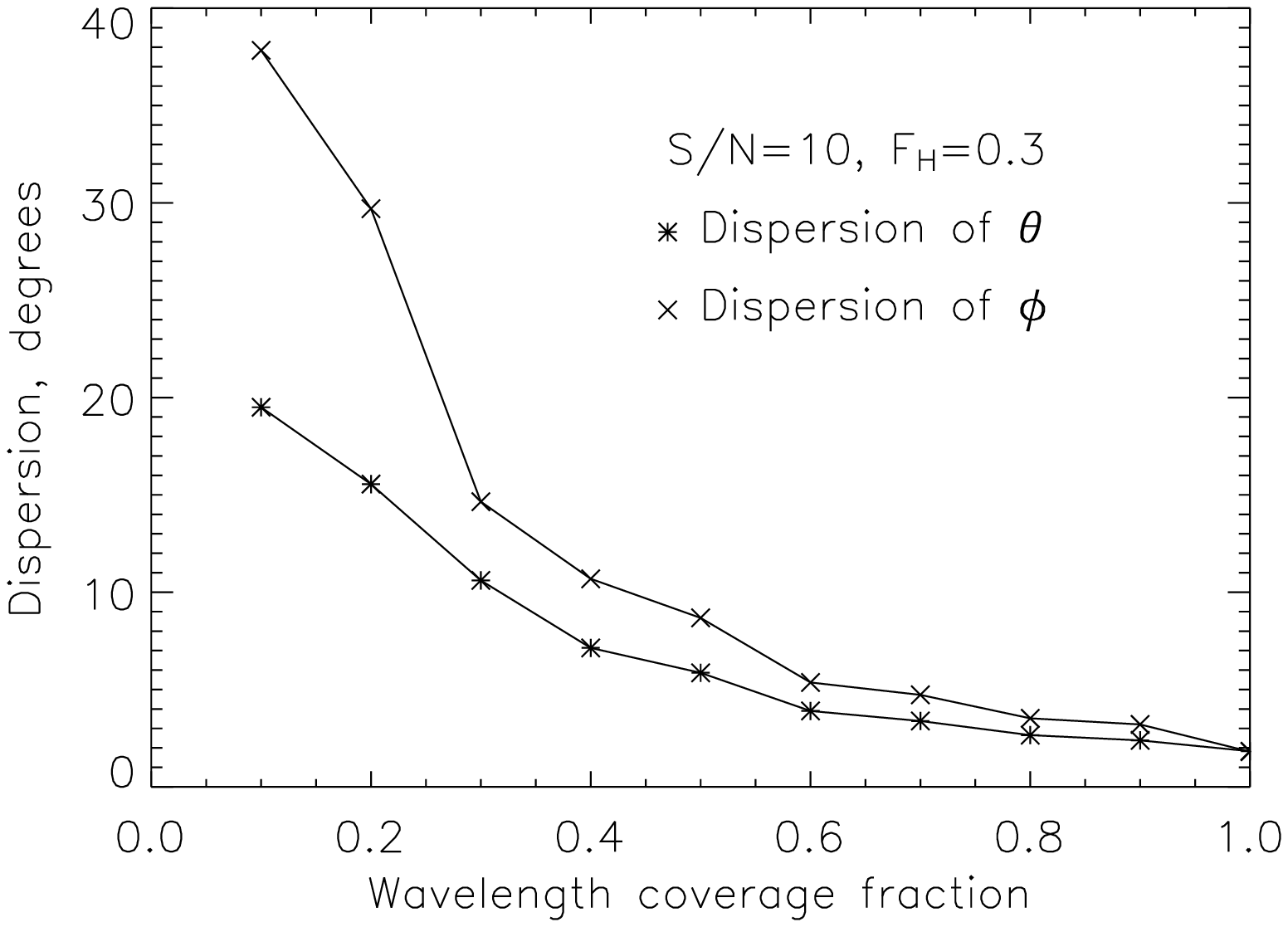}
  \caption{The rms dispersion, in degrees, on the measurement of
    $\phi_{H}$ and $\theta_{H}$ for simulated spectra with varying
    fractions of eigenspectrum wavelength coverage.  The host galaxy
    luminosity fraction and $S/N$ ratio were held constant, and
    noise was added to the template spectra according to the $S/N$
    level. \label{F_waverange}}
\end{figure}

Figure\,\ref{F_waverange} shows the rms dispersion of the galaxy
classification angles measured for a large number of simulated
spectra, as a function of the fraction of the total eigenspectrum
wavelength covered.  The host galaxy fraction and $S/N$ level were
held constant in all of the simulations.  As expected, the dispersions
increase with shorter wavelength ranges.  For our purposes, the
dispersions become unacceptably large at a wavelength coverage fraction
of about 0.5.  At that point, the dispersions are comparable to those
in cases with low $S/N$ and small galaxy fraction.  In SDSS spectra,
the redshift corresponding to a coverage fraction of 0.5 is
$z_{limit}=0.752$, which is what we adopt as the upper limit  
for the sample selection.

The redshift limitation is not a limitation of the technique in general,
but applies to the specific case of the eigenspectra used here.  It is
also restricted to the galaxy eigenspectra, since the quasar eigenspectra
cover a much wider wavelength range.  The limitation could be overcome
by constructing galaxy eigenspectra that cover a wider wavelength
range, especially at shorter wavelengths.  This requires UV spectra of
low-redshift galaxies, or optical spectra of galaxies at higher redshifts.

%
\begin{figure*}
  \plotone{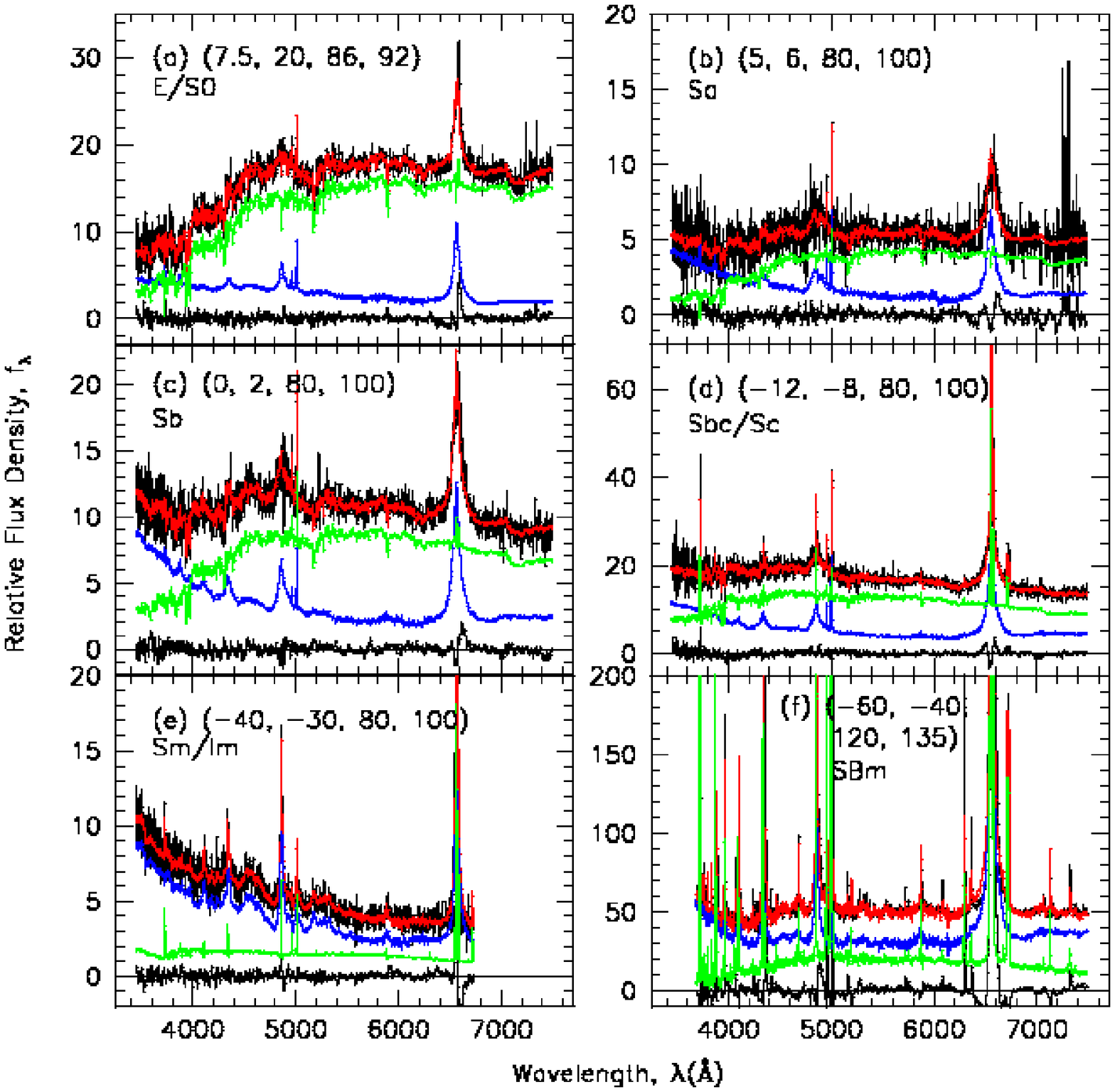}
  \caption{Examples of eigenspectrum reconstructions of AGN/host galaxy
    spectra.  Each panel shows the original spectrum (black), the
    reconstructed spectrum (red), the residual spectrum smoothed by 7
    pixels (black, near $f_{\lambda}=0$), the AGN component (blue), and
    the host galaxy component (green).  The six examples were drawn from
    the $\phi_{H}$ and $\theta_{H}$ regions, defined by the coordinates
    inside panel, from which the galaxy template spectra were constructed.
    The estimated morphological type of galaxies inside each region
    are given.  The emission lines of the spectra in the lower-right
    panel extend well-beyond the top limit of the plot.  \label{F_specEx}}
\end{figure*}

\subsection{Example Reconstructions \label{S_exfits}}
Examples of fits to real spectra are shown in Fig.\,\ref{F_specEx}.
The spectra were drawn from the SDSS dataset (see \S\,2),
and were selected from each of the six regions for which the template
galaxy spectra were created.  The $S/N$ ranges from 10 to 50 and
the host galaxy fractions vary from $0.25$ to $0.71$.  The original
spectrum, the eigenspectrum reconstruction, the residual spectrum the
AGN component, and the host galaxy component are shown for each object.
It is evident from the figures that the spectra are well fit by the
combined eigenspectrum sets.  Larger residuals occur in the regions
containing strong narrow emission lines, as expected, because a larger
number of eigenspectra is often required to describe narrow line features.

%
%
\section{SDSS Host Galaxy and AGN Spectra}
  \label{S_spec}

\subsection{Decomposition of SDSS AGN Spectra \label{S_sdssDecomp}}

%
\begin{figure}
  \plotone{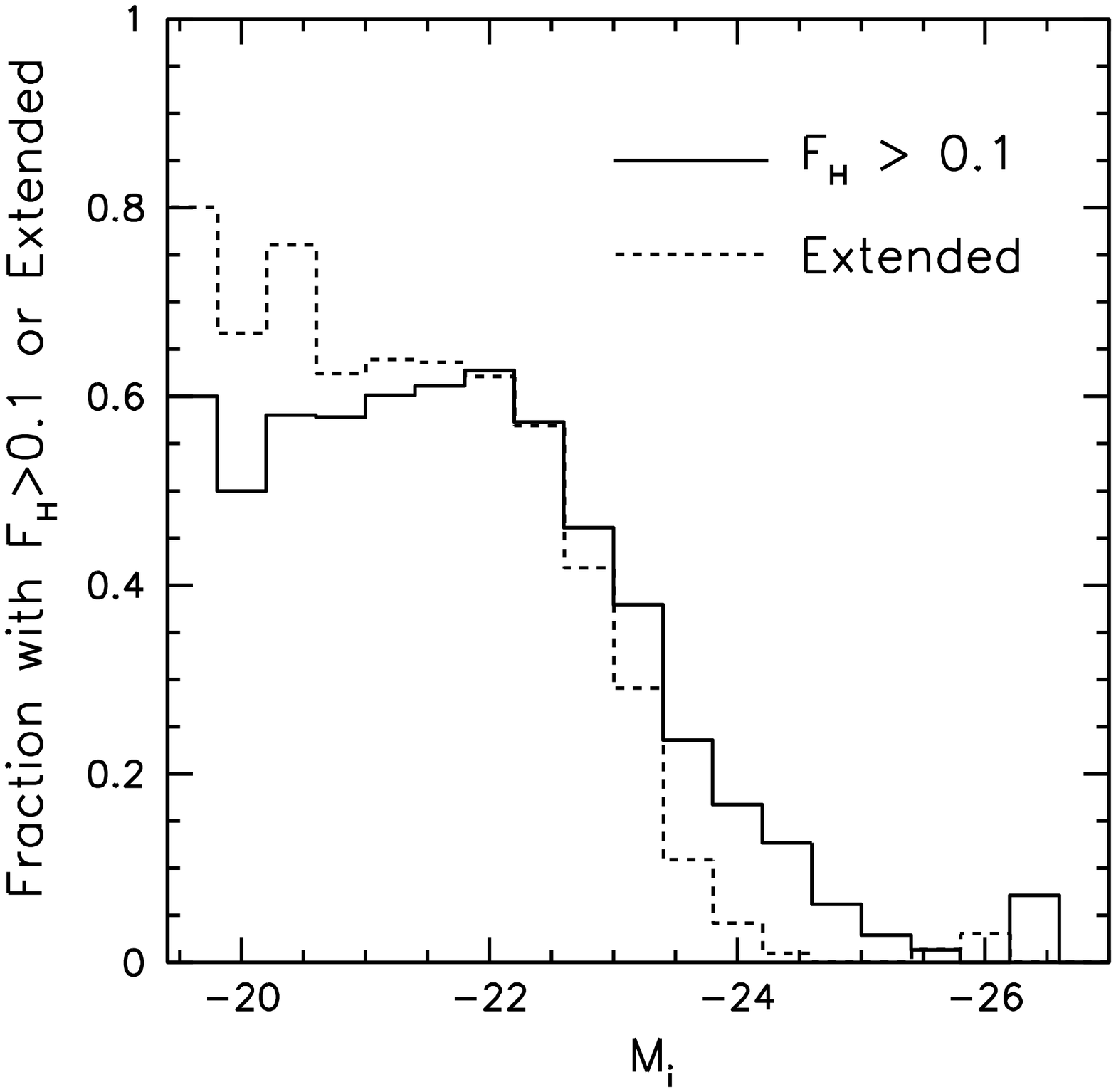}
  \caption{The fraction of AGN spectra with $S/N>10$ having a detectable
    host galaxy, i.e.\ a galaxy fraction greater than $F_{H}=0.1$, as a
    function of absolute $i$ band cmodel magnitude, $M_{i}$.  Also shown is
    the fraction of AGNs in the sample that have extended image morphology
    as a function of $M_{i}$.
    \label{F_detectFrac}}
\end{figure}

The eigenspectrum decomposition method was applied to all $11,647$ AGN
spectra in the initial sample.  The fits to some of the spectra resulted
in components that extend below zero flux density in continuum regions.
Inspection of those cases showed that the AGN component dominates enough
that the galaxy component is too small to be reliably estimated, that is,
the galaxy fraction is usually well below $0.1$ --- the practical limit
discussed in \S\,3.4.  Fits to spectra with $S/N$ ratios
less than about 10 are also unreliable (\S\,3.4), so those
spectra are removed for further analysis.  In Fig.\,\ref{F_detectFrac} the
fraction of AGNs with $S/N>10$ having a host galaxy fraction exceeding $F_{H}
= 0.1$ is shown as a function of the AGN absolute $i$ band magnitude.
As expected, the fraction of AGNs with detectable host galaxy components
drops with increasing luminosity.  The detected fraction falls with
luminosity in much the same way as the fraction with extended profiles
in the imaging data (Fig.\,\ref{F_detectFrac}).  There is a close
correspondence between the classification of an AGN image as extended,
and the AGN having a spectroscopic host fraction $F_{H}>0.1$.  Of the
AGN with $z<0.752$, $52\%$ are extended, and $61\%$ have $F_{H}>0.1$.
Of the AGN with extended images, $89\%$ have $F_{H}>0.1$ compared with
only $30\%$ in the AGN sample with unresolved images.

%
\begin{figure}
  \plotone{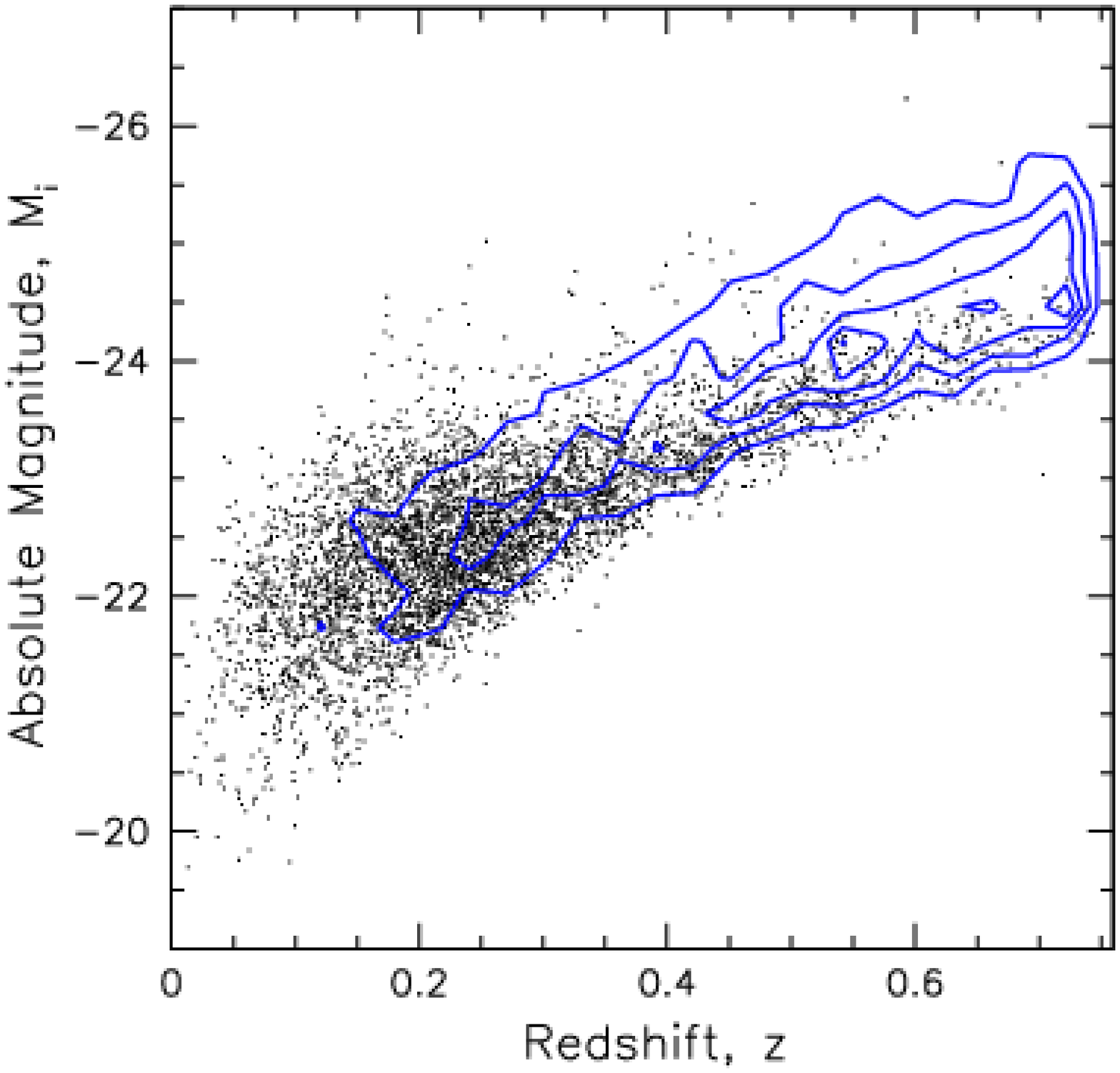}
  \caption{The absolute $i$ band cmodel magnitude, $M_{i}$, vs. redshift for
    the SDSS AGN sample with $S/N>10$, up to a redshift of $z=0.752$. Dots
    represent AGNs for which a host galaxy was reliably detected;
    that is the spectra have a host galaxy fraction $F_{H}>0.1$.
    Contours represent the region occupied by AGNs without a reliably
    detected host galaxy. There are $7664$ AGNs in the $S/N>10$ sample,
    $4666$ of which have a detected host galaxy.  The AGNs with detectable
    hosts are generally at lower luminosities and redshifts.
    \label{F_zedMi}}
\end{figure}

Figure~\ref{F_zedMi} shows the $i$ band cmodel (total flux) absolute
magnitude vs.\ redshift for the initial sample of $7664$ SDSS AGN
with $S/N>10$; squares represent the 4666 AGNs for which there is a
``reliably'' detected host galaxy.  At redshifts greater than about $0.4$,
the hosts of less luminous AGNs are often undetected because the $S/N$
is low, and they are undetected in high luminosity AGNs because the host
component fraction is too low.  For most of the analysis that follows,
we will omit AGNs with a host component fraction less than $F_{H} = 0.1$,
or a median spectroscopic $S/N$ per pixel in the $i$ band wavelength
region of less than 10.

%
\begin{figure}
  \plotone{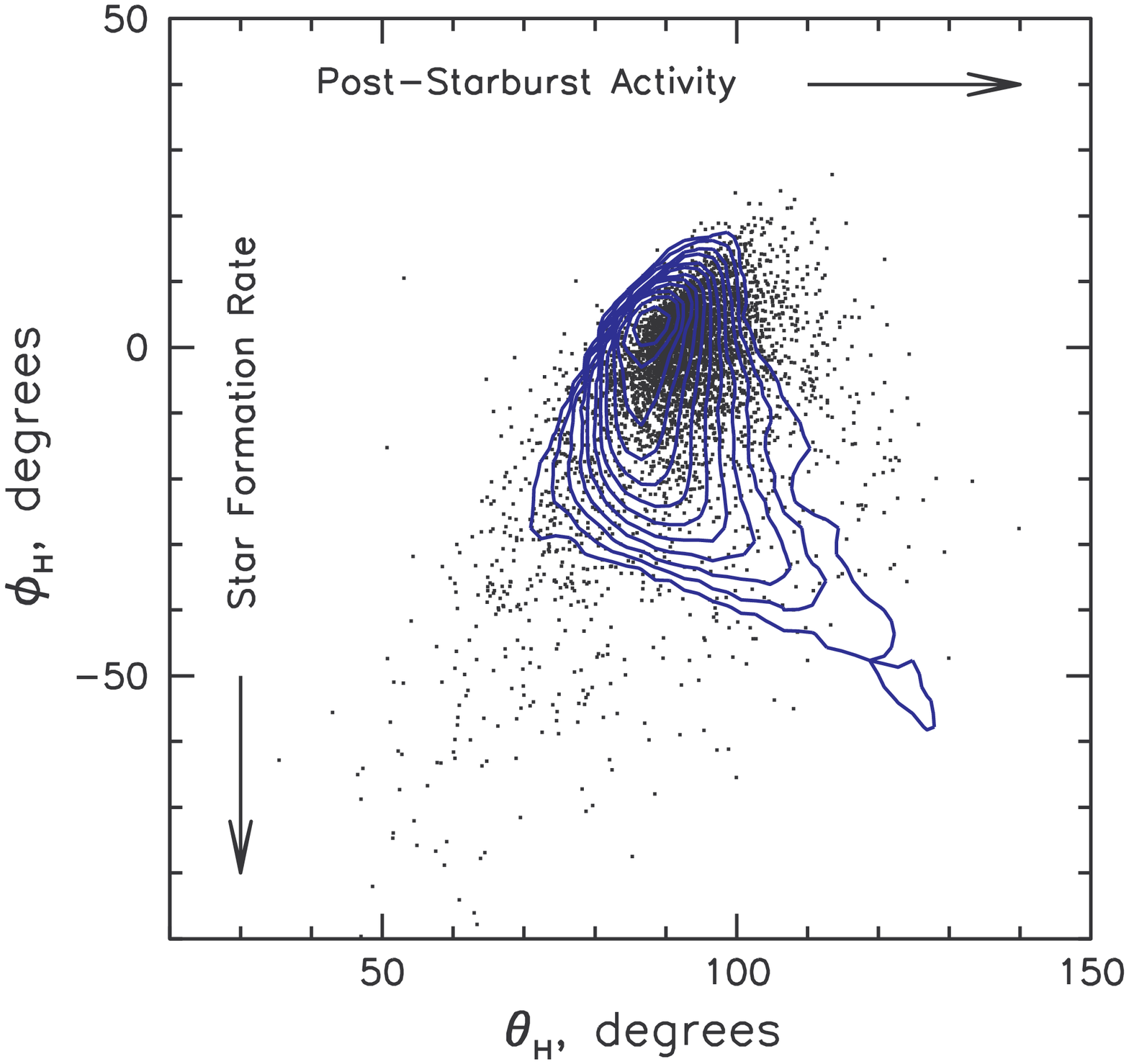}
  \caption{The galaxy classification angles (points), $\phi_{H}$ and
    $\theta_{H}$, found for SDSS broad-line AGN spectra with $z<0.752$,
    $S/N>10$, and a host galaxy fraction $F_{H}>0.1$.  There are $4666$
    AGNs in the sample.  The contours show the distribution of the
    classification angles for non-active galaxies from the sample
    described by \citet{yip04a}; each contour level differs by a factor
    of two relative to adjacent levels.  Labels show the general
    trends of star formation rate and post-starburst activity with
    the classification angles. \label{F_phiThetaH}}
\end{figure}

The galaxy classification angles for the fits to the 4666 ``reliably''
detected host galaxies are shown in Fig.\,\ref{F_phiThetaH}.  The
distribution of host galaxy angles is clustered in a relatively small
region, as are those of the normal galaxies analyzed by \citet{yip04a},
which are shown with density contours in the figure.  Labels show the
general trends of star formation and post-starburst activity with the
classification angles; however, the labels are meant only to suggest the
correlations, but not that there is a strict one-to-one correspondence.
A more detailed comparison of the distribution relative to normal galaxies
is discussed in \S\,5.

Information for the objects used in this study is tabulated in
an electronic table, the column description for which is given
Table\,\ref{tab1}.  The information is described in the electronic
version of Table\,\ref{tab2}, which includes all (11,647 total) SDSS
broad-line AGNs with redshifts $z<0.752$ for which an eigenspectrum
reconstruction was attempted.  The table includes the SDSS coordinate
name, coordinates, redshift, plate number, fiber number, spectroscopic
MJD, imaging morphology, spectroscopic $S/N$ in the $i$ band, magnitudes,
luminosities (described in \S\,4.2), host galaxy
fraction ($F_{H}$), and the AGN and host galaxy classification angles.
Default values (set to zeros) of the reconstructed spectral parameters
are given if either the quasar or host galaxy reconstructed spectra
had negative flux densities in the $4160-4210${\AA} continuum region.
The objects are listed regardless of $S/N$ or host galaxy fraction,
$F_{H}$; users of the dataset can therefore apply any range of selection
cuts to the sample.

\subsection{Luminosity Determination \label{S_correctFiber}}

Estimates of the luminosities of the host galaxy and AGN components were
made by measuring rest frame absolute magnitudes in the SDSS $g$ and $r$
passbands.  Normally, absolute magnitudes are determined by measuring
the apparent magnitude of an object in an observed frame passband,
then applying the distance modulus (using a specific cosmology)
and a K-correction, which accounts for the difference between the
observed frame and rest frame flux density covered by a given passband.
The wavelength range of the eigenspectra cover the SDSS $g$ and $r$
bands completely, so synthetic magnitudes can be constructed without the
need for K-corrections.  The selection of the SDSS $g$ and $r$ filters
also allows for direct comparison between the results here and those
of other SDSS galaxy studies, such as the ones by \citet{bernardi03c,
bernardi03b} and \citet{hogg04}.

The rest frame apparent magnitudes were measured directly from the
reconstructed spectra by convolving the spectra with the SDSS filter
transmission curves\footnote{http://www.sdss.org/dr3/instruments/imager/},
including 1.3 airmasses of extinction, after shifting the spectra to the
rest frame and multiplying the flux densities by the redshift expansion
factor, $1+z$.  The distance modulus (as a function of redshift and the
selected cosmology) was applied to the apparent magnitudes to determine
the absolute magnitudes; K-corrections are not necessary.  The absolute
magnitudes of the reconstructed AGN component, the reconstructed host
galaxy component, and the total spectrum of each object were determined
using the spectroscopic convolution technique.

The determination of the absolute magnitudes using the spectroscopic
convolution technique is possible due to the high quality of the SDSS
spectrophotometry\footnote{http://www.sdss.org/dr3/products/spectra/spectrophotometry.html}.
The airmasses of the spectroscopic observations vary from plate to plate,
so slight improvements could perhaps be made by adjusting the airmass
of the extinction applied to the transmission curves.  Additionally,
small uncertainties could be introduced if fibers are not precisely
centered on the active nuclei.  However, a direct comparison between the
stellar imaging and spectroscopic magnitudes for tens of thousands of
stars in the SDSS DR3 sample, shows an rms variation of less than 0.06
magnitudes in the $g$, $r$, and $i$ bands at a spectroscopic $S/N$ of 10.
That is significantly less than the uncertainty expected to be introduced
by quasar variability.  We expect the spectrophotometric accuracy to be
somewhat lower than $0.06$ magnitudes for the AGN and host components
separately, due to the uncertainties of the eigenspectrum reconstruction
technique (\S\,3).

Before using the magnitudes to study the properties of the objects,
a correction must be applied to account for the finite angular width of
the SDSS spectroscopic fibers, which subtend only a $3\arcsec$ diameter
on the sky.  Not all of the collected light from an active nucleus
or host galaxy will be covered by a fiber, and the fraction of the
total intensity entering the fibers will depend on the sizes of the images.
An AGN component is spatially unresolved, while a host galaxy may be
extended, so the correction factor applied to account for the finite
fiber diameter will be different for each component.  In addition,
varying seeing conditions can affect the amount of light entering the
spectroscopic fibers, even from unresolved sources.

The offsets needed to correct the (unresolved) AGN spectroscopic
magnitudes were determined by comparing the spectroscopic magnitudes of
non-variable early-type stars observed in the SDSS spectroscopic survey
to their so-called ``cmodel'' magnitudes.  The cmodel magnitudes $m_{c}$
\citep{abazajian04}, which are derived from model profile fits to the
images, are designed to include nearly all of the flux of both point
sources and extended galaxies.  The mean observed frame magnitude offsets
in the $g$, $r$, and $i$ bands were determined for each spectroscopic
plate separately, to account for varying observing conditions.
Each plate corresponds to observations of about 25 usable stars.
The mean offsets were found by weighting each star by the square of its
spectroscopic $S/N$, although the weights did not usually significantly
affect the results.

In the analysis that follows, the magnitude offsets $\Delta g_{A}$ and
$\Delta r_{A}$, were applied to each set of rest frame AGN absolute
$g$ and $r$ magnitudes.  Because the $g$, $r$, and $i$ offsets were
determined in the observed frame, the rest frame $g$ and $r$ corrections
applied to the AGN magnitudes were interpolated from the two closest
observed frame offsets, at the redshifted central wavelengths of the
$g$ and $r$ bands.  Interpolation caused only minor corrections to the
magnitude offsets, because the mean differences among the offsets of the
three passbands are only about $0.02$ magnitudes.  This is much smaller
than the uncertainties between the spectroscopic and imaging epochs
introduced by AGN variability, which are typically a few tenths of a
magnitude \citep{vandenberk04}, and become relevant when correcting the
galaxy magnitudes.  There are contributions to the uncertainty in the AGN
magnitude corrections from photometry, the spectroscopic decomposition,
and varying observing conditions.  The rms dispersion of the spectroscopic
to cmodel magnitudes for stars (noted above), places a lower limit on
the uncertainty of $0.06$ magnitudes.  We estimate the typical total
uncertainty to be up to a few tenths of a magnitude.  The corrected AGN
absolute magnitudes $M_{g,A}$ and $M_{r,A}$, and the $\Delta g_{A}$
and $\Delta r_{A}$ values applied to the AGN spectroscopic absolute
magnitudes are given in the electronic version of Table\,\ref{tab2}.
Default values (set to zeros) are given for the quantities if the
spectroscopic decomposition was non-physical (see \S\,4.1).

A significantly larger fraction of the host flux is expected to fall
outside the fiber area, compared to the AGN flux.  The same method used to
correct the AGN flux cannot be used to correct the host galaxy component
flux, because host galaxy sizes and profiles can vary greatly from object
to object.  Instead, we estimate the total host galaxy flux by subtracting
the corrected AGN flux from the total photometric cmodel flux of the AGN
plus host galaxy.  In practice, the magnitude offset for a host galaxy,
$\Delta m_{H}$, was found by estimating the ratio of the cmodel (subscript
$c$) and spectroscopic (subscript $s$) fluxes of the host galaxy
\begin{eqnarray}
  \Delta m_{H} & = & m_{c,H} - m_{s,H} \\
               & = & -2.5\log(f_{c,H}/f_{s,H})\,.
  \label{Eq_dmH}
\end{eqnarray}
The host galaxy cmodel to spectroscopic flux ratio is a measure of
the amount of host galaxy flux lost outside a spectroscopic fiber.
The ratio cannot be measured directly, because of the contribution of
the AGN component, but it can be estimated by subtracting the estimate
of the AGN flux
\begin{eqnarray}
  \frac{f_{c,H}}{f_{s,H}} & = & \frac{1}{f_{s,H}}(f_{c,T} - f_{c,A}) \\
    & = & \frac{1}{f_{s,H}}(S_{T}f_{s,T} - S_{A}f_{s,A}) \\
    & = & S_{T}{F}^{-1}_{m,H} - S_{A}(F^{-1}_{m,H} - 1)\,,
  \label{Eq_fHratio}
\end{eqnarray}
where the subscript $T$ refers to the total of the AGN (subscript $A$)
and host (subscript $H$) quantities, and the cmodel to spectroscopic
flux ratios are
\begin{eqnarray}
  S_{T} & = & f_{c,T}/f_{s,T} \\
  S_{A} & = & f_{c,A}/f_{s,A}\,.
  \label{Eq_STSA}
\end{eqnarray}
The quantity $F_{m,H}$ is the spectroscopic host contribution fraction, similar
to $F_{H}$ defined by Eq.\,\ref{Eq_FH}, but using the flux in the rest frame
$g$ or $r$ bands.  The result for the host galaxy magnitude correction is
\begin{eqnarray}
  \Delta m_{H} & = & -2.5\log(S_{T}F^{-1}_{m,H} - S_{A}(F^{-1}_{m,H} - 1))\,.
  \label{Eq_dmHfinal}
\end{eqnarray}

The magnitude offsets were applied to the host galaxy absolute magnitudes
in the analysis of the following sections.  The cmodel to spectroscopic
flux ratios, $S_{A}$ and $S_{T}$, can only be determined in the observed
frame, so the values used in Eq.\,\ref{Eq_dmHfinal} were interpolated
from the observed frame to the redshifted wavelengths of the $g$ and
$r$ bands.  While the corrected magnitudes should give an accurate
estimate of the luminosities for the sample as a whole, the values for
individual cases may be less reliable due to measurement uncertainties,
unaccounted for differences in observing conditions, AGN variability,
and inaccuracies in the spectral reconstruction.  The typical uncertainty
from those sources is estimated to be a few tenths of a magnitude, but
larger than the uncertainty for the AGN measurements.  In a small number of
cases, the ``corrections'' make the total host luminosities fainter than
the spectroscopic luminosities. The result of the uncertainties is to
broaden the distribution of luminosities.  In any case, the corrected
host galaxy absolute magnitudes are listed in the electronic version
of Table\,\ref{tab2}, along with the magnitude offsets $\Delta g_{H}$
and $\Delta r_{H}$ that were applied to the spectroscopic magnitudes,
so that the results may be compared between the spectroscopic and
corrected luminosities.  Default values (set to zeros) are given
for the quantities if the spectroscopic decomposition was non-physical
(see \S\,4.1).

\section{Results \label{S_results}}
%
%

%
\begin{figure*}
  \plottwo{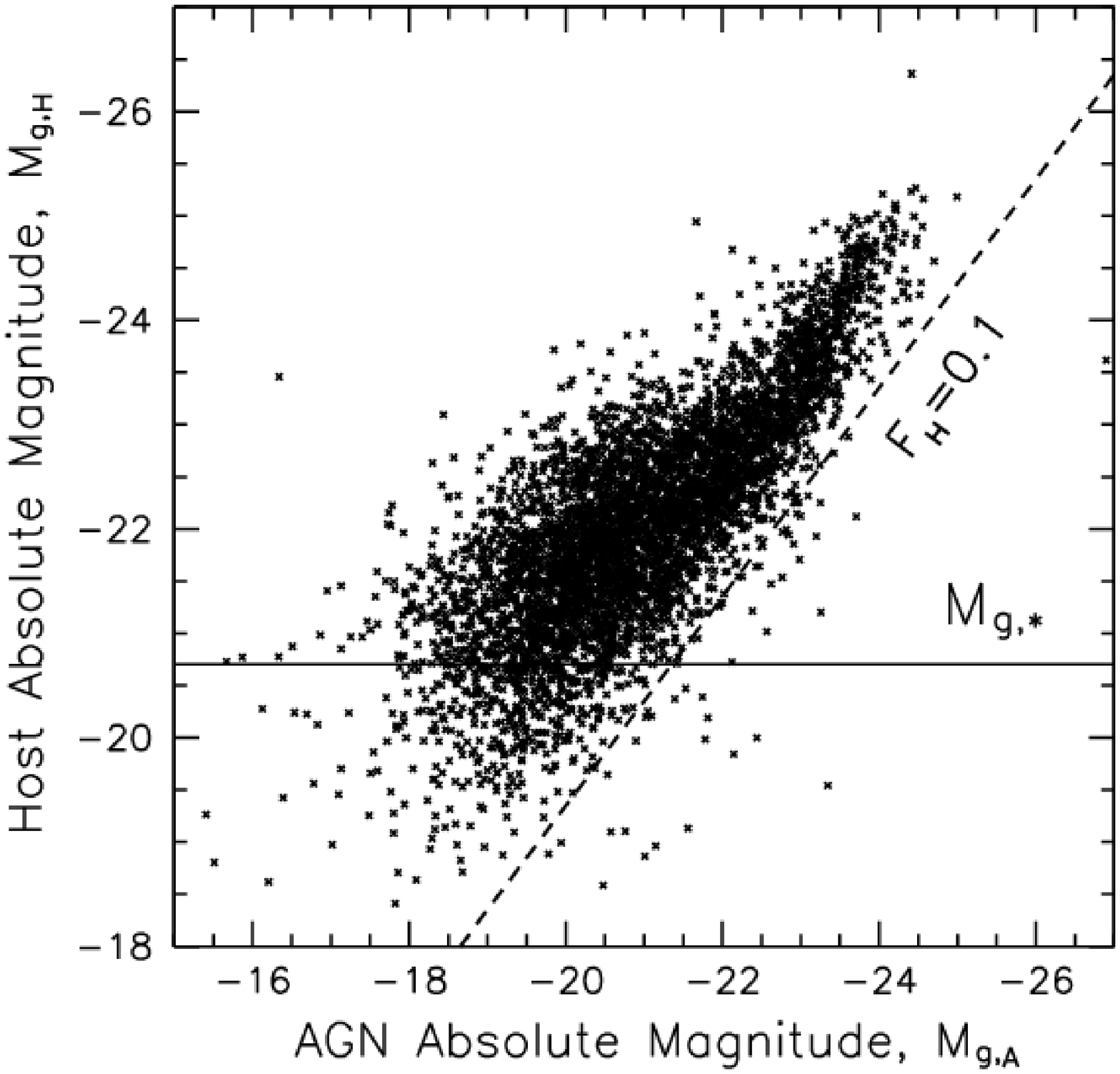}{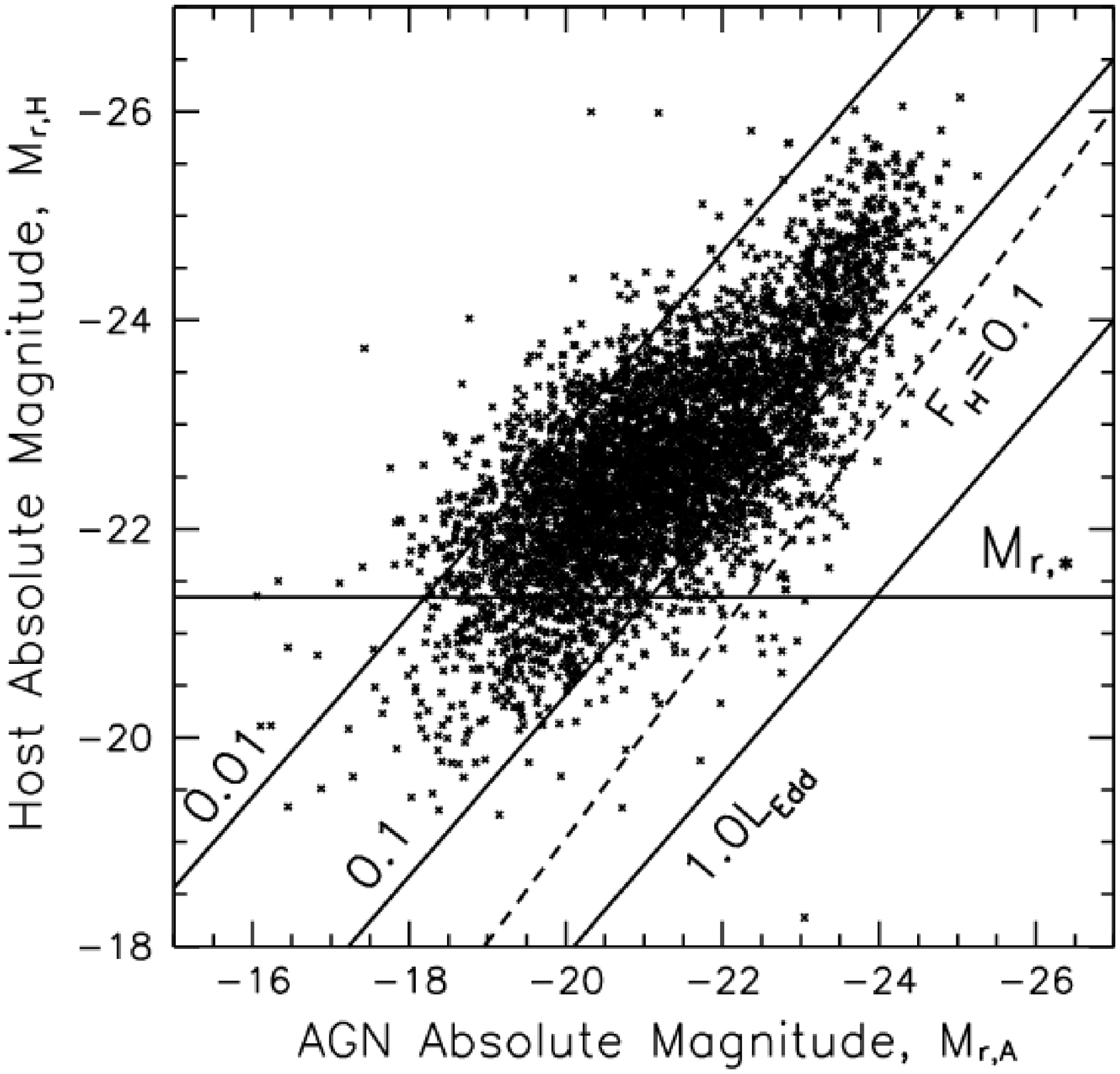}
  \caption{Absolute $g$ band (left) and $r$ band (right) magnitudes of
    the AGN and host galaxy components, $M_{H}$ and $M_{A}$
    respectively. The dashed diagonal lines marked $F_{H}=0.1$ correspond
    to a host galaxy fraction of 0.1 in each band, adjusted for the
    average correction from spectroscopic to cmodel magnitudes,
    and closely mark the sensitivity limits for the detection of
    host galaxies.  The horizontal lines show the evolution corrected
    characteristic magnitude of early type galaxies without broad-line
    AGNs, $M_{*}$.  The solid diagonal lines in the $r$ band plot show
    the host luminosity at which an AGN would be expected to radiate at
    $0.01, 0.1$, and $1.0$ times the Eddington luminosity, $L_{Edd}$.
    \label{F_MgrAH}}
\end{figure*}

\subsection{Comparison of AGN and Host Galaxy Luminosities\label{S_MAH}}
It was shown in \S\,4.1 that the fraction of AGNs
with a detectable host galaxy component ($F_{H}>0.1$) decreases with
AGN luminosity.  Fig.\,\ref{F_MgrAH} shows the host galaxy absolute
magnitudes vs.\ the AGN absolute magnitudes for all of the objects with a
spectroscopic $S/N > 10$.  The AGN and galaxy luminosities appear to be
closely correlated in both the $g$ and $r$ bands.  This is due in part
to the increased difficulty of detecting a host galaxy as its fractional
contribution to the spectra decreases.  The dashed line shows where the
spectroscopic host luminosity contributes $10\%$ to the total luminosity,
adjusted for the average magnitude changes to the AGN and host components
(\S\,4.2), and corresponds closely to the galaxy
fraction reliability limit, $F_{H}=0.1$ discussed in \S\,3.4.
The selection function, that is, the fraction of spectra in which a
host galaxy could be detected, decreases toward that limit.  Points may
extend beyond the limit (and a few do), depending upon the spectroscopic
to cmodel magnitude corrections described in \S\,4.2.
A correlation between the host and AGN luminosities cannot be established
without a more careful analysis of the selection function (which we
defer to future work).  However, if the (unplotted) AGNs with hosts
outside of the detection limits radiate at less than the Eddington
limit (see below), there would be a very strong correlation between
the AGN and host galaxy luminosities.

Also shown in Fig.\,\ref{F_MgrAH} are the evolution corrected
characteristic absolute magnitudes for early type galaxies (without broad
emission line components), $M_{g,*} = -20.70$ and $M_{r,*} = -21.35$,
as found by \citet{bernardi03c} for the median redshift of our sample,
$\langle z \rangle = 0.236$.  The median values of the host absolute
magnitudes are $\langle M_{g,H} \rangle = -22.03$ and $\langle M_{r,H}
\rangle = -22.62$, which are clearly more luminous than typical elliptical
galaxies.  There is also a wide distribution in the host luminosities
at every AGN luminosity.

A number of image decomposition studies have found that there is a lower
threshold to the galaxy luminosity for a given AGN luminosity. This
host--AGN luminosity threshold is thought to arise from scaling
relationships among bulge luminosity, bulge mass, black hole mass, and
Eddington luminosity.  For example, \citet{dunlop03} found that quasars
are typically radiating at $1$--$10\%$ of their Eddington luminosities,
which in effect establishes a scaling relation between AGN luminosity and
black hole mass.  As there is also a correlation between black hole mass
and bulge mass and luminosity, one expects to find a scaling relation
between AGN and host galaxy luminosity.  Thus, a given AGN luminosity
requires a host galaxy that is sufficiently massive and luminous to
harbor a black hole of the mass inferred from the Eddington ratio.

We have plotted the expected relationships between AGN and host
luminosities for the $r$ band in Fig.\,\ref{F_MgrAH}, for Eddington
luminosity ratios $L_{A}/L_{Edd}=1.0, 0.1$, and $0.01$.  The relationship
between galaxy luminosity and black hole mass was estimated by assuming
the $r$ band bulge mass-to-light ratio found by \citet{bernardi03c}, a
black hole mass to bulge mass of $0.0012$ \citep[e.g.][]{mclure02}, and
assuming that the host luminosity is due entirely to a bulge component.
The relationship between black hole mass and AGN luminosity was estimated
by assuming a $10\%$ accretion radiation efficiency for the bolometric
Eddington luminosity, and using the $V$ band to bolometric correction
factor given by \citet{elvis94} (which should be close to the $r$ band
factor within the uncertainties).  Equating the black hole masses inferred
from the galaxy luminosity and AGN luminosity gives the relationship
between the two luminosities as a function of the Eddington ratio.
For most spectra, the sensitivity limits of our study do not extend to an
Eddington ratio of unity, but only to $L_{A}/L_{Edd}$ of a few tenths.
The vast majority of our AGNs appear to be radiating at greater than
$0.01 L_{Edd}$.  It is possible that the sample is missing low-luminosity
AGNs with broad emission lines that are too weak to be detected in the
SDSS spectra.  Those objects could have $L_{A}/L_{Edd}$ much smaller
than $0.01$, and therefore could populate the upper left regions of
the plots in Fig.\,\ref{F_MgrAH}.  Until such objects can be
confirmed or ruled out, we cannot conclude that there is a lower limit
to $L_{A}/L_{Edd}$.

Because we have spectroscopic information for the AGN components, it
will be possible in many cases to derive estimates for the black hole
masses based on the AGN broad emission line widths and the relationship
between broad emission line region size and luminosity; we leave these
issues to be addressed in future work.  In any case, we can conclude that
AGNs may reside in hosts with a wide range of luminosities, but that in
general the hosts of AGNs with $M \lesssim -18$ are significantly more
luminous than galaxies without strong broad line AGNs.

\subsection{Host Galaxy Colors\label{S_hostColors}}

%
\begin{figure}
  \plotone{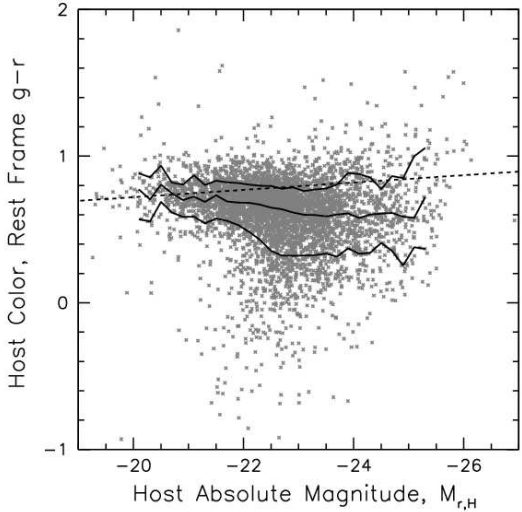}
  \caption{Rest frame host galaxy color vs.\ host galaxy absolute
    magnitude in the $r$ band, $M_{r,H}$.  The median value of the
    color in bins of 0.2 magnitudes is shown along with the $68.3\%$
    confidence limits.  The dashed line shows the color-absolute
    magnitude relation for early type galaxies found by \citet{bernardi03c}.
    \label{F_hostColor}}
\end{figure}

The host galaxy rest frame ($g-r$) color is shown as a function
of host absolute magnitude, $M_{r,H}$, in Fig.\,\ref{F_hostColor}.
For clarity, the mean colors and $68.3\%$ confidence envelope in bins 0.2
magnitudes wide are also shown, connected by lines.  There appears to
be a negative correlation of host color with luminosity, such that the
hosts become bluer at higher luminosities.  There is also a fairly wide
distribution of colors at any given luminosity.  Aside from measurement
uncertainties, it is possible that the the wide distribution of colors
and the luminosity trend are due to a wide distribution of galaxy types
that changes with luminosity, because the colors of galaxies differ by
type \citep[cf.][]{hogg04}.  The straight line in Fig.\,\ref{F_hostColor}
shows the color-luminosity relation for elliptical galaxies found by
\citet{bernardi03a}, which is similar to the relation for bulge-dominated
galaxies found by \citet{hogg04}.  The range of host galaxy colors is
much wider than the dispersion around the elliptical color-luminosity
relation, which is less than $0.1$ magnitude \citep{bernardi03a}.
The color line for elliptical galaxies describes the host galaxy colors
reasonably well at low luminosities ($M_{r,H}>-21.5$), but it becomes
increasingly redder than the majority of the hosts at high luminosity.
This implies that if host galaxies are mainly ellipticals or otherwise
bulge-dominated, for a given luminosity they are bluer than the same
types of galaxies that do not contain an observed AGN.  The results
may instead mean that the types of galaxies that contain AGNs change
with luminosity, such that disk-dominated hosts become more common
at higher luminosity.  However, not only do most image decomposition
studies find that the majority of host galaxies are bulge-dominated,
but it also appears that bulge-dominated hosts are more common at higher
luminosities \citep{dunlop03}.  Multi-band photometric imaging studies
also support the idea that the host galaxies are bluer than their
non-active counterparts \citep{hutchings02, sanchez04, jahnke04a}.

\citet{bernardi03a} showed that the color luminosity relation for
early type galaxies is a consequence of a strong dependence of color on
stellar velocity dispersion.  Based on this, we would expect the redder
hosts in our sample at a given luminosity to have a higher velocity
dispersion, and consequently a larger black hole mass.  However, if the
star formation rate is related to the presence of an AGN, that prediction
may not hold.  There is evidence from the eigencoefficients, described
in \S\,5.3, that the hosts are dominated by post-starburst
galaxies, which would likely change both the color-velocity dispersion
and color-luminosity relation from those of non-active ellipticals.
Further analysis in the future of the spectroscopic components to
estimate velocity dispersion and black hole mass should help shed light
on these issues.

The nuclear colors of an AGN are expected to change through time in a
number of possible evolutionary sequences.  For example \citet{storchi01}
describe a scenario in which a merger event would both fuel AGN activity,
and trigger circumnuclear star formation early in the process.  Dust
associated with the star formation may reprocess the nuclear light, making
it appear red, so that in this early phase the galaxy would be classified
as an ultraluminous infrared galaxy \citep[e.g.][]{cid01,bushouse02}.
With time, dust is removed and the starburst fades, so that an unobscured
AGN combined with an older stellar population dominate the nucleus.
The objects in our study may be in such a later evolutionary phase,
as long as the aged nuclear stellar population is younger than it would
have been in the absence of the triggering event.

\subsection{Host Galaxy Spectroscopic Classification\label{S_specClass}}

%
\begin{figure}
  \plotone{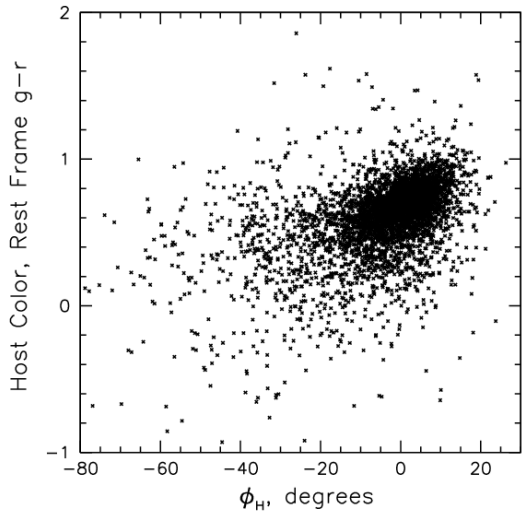}
  \caption{Host galaxy rest frame $g - r$ color vs.\ $\phi_{H}$ angle
    values for spectra with $S/N>10$ and $F_{H}>0.1$.
    \label{F_phiColorHost}}
\end{figure}

Further information about the classification of the host galaxies
and the AGNs is contained in the eigencoefficients.  The relative
strengths of the eigencoefficients contain information about the
physical properties of the galaxy and AGN components.  While there has
yet been no quantitative calibration of the eigenspectra with physical
parameters, there are clear qualitative correlations.  The first few
AGN eigenspectra are related to the so-called ``eigenvectors 1 and 2''
in observed parameter space as defined by \citet{boroson92}, and it is
believed that the eigenvectors are indicators of black hole mass and
accretion rate \citep{boroson02}.  The galaxy classification angles,
$\phi_{H}$ and $\theta_{H}$, are related to star formation rate and
the presence of post-starburst activity respectively \citep{yip04a,
connolly95,castander01}.  Figure\,\ref{F_phiColorHost} shows how the $g-r$
host galaxy colors, discussed in \S\S\,5.1 \& 5.2,
are related to the first galaxy classification angle, $\phi_{H}$.
The angle $\phi_{H}$ and the host color are correlated, as expected given
the qualitative relation between $\phi_{H}$ and star formation rates.

%
\begin{figure}
  \plotone{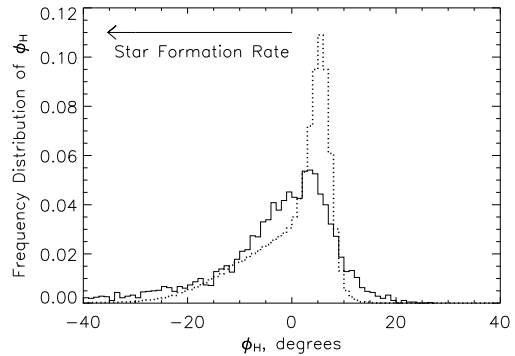}
  \caption{Distribution of the host galaxy $\phi_{H}$ angle values for
    spectra with $S/N>10$ and $F_{H}>0.1$ (solid histogram).  The dashed
    histogram shows the $\phi_{H}$ distribution for normal galaxies,
    found by \citet{yip04a}.
    \label{F_phiHist}}
\end{figure}

Fig.\,\ref{F_phiHist} shows the distribution of $\phi_{H}$ for all of
the reliably detected host galaxies with $F_{H}>0.1$ and spectroscopic
$S/N>10$.  Also shown is the $\phi_{H}$ distribution for the normal
galaxies analyzed by \citet{yip04a}.  There are clear differences in the
distributions.  The host galaxy distribution is shifted to slightly more
negative (bluer color) values, and there is a much less pronounced (and
blueward shifted) narrow peak at $\phi_{H}\approx 5\arcdeg$ compared to
the inactive galaxies.  The distribution of the host $\phi_{H}$ values
probably indicates that the host star formation rates are generally
higher than for galaxies without AGNs.  That would be true especially if
the morphology distribution of the hosts relative to other galaxies is
weighted toward ellipticals, as indicated in a number of imaging studies
\citep[e.g.][]{jahnke04a, sanchez04, hamilton02, percival01, schade00}.

%
\begin{figure}
  \plotone{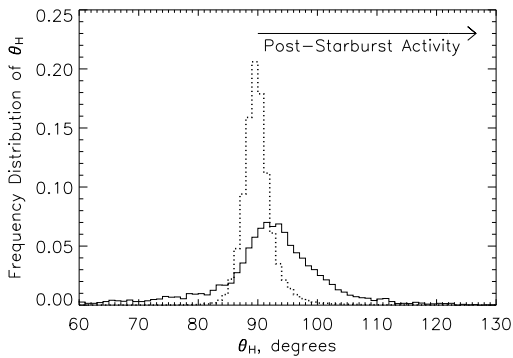}
  \caption{Distribution of the host galaxy $\theta_{H}$ angle values for
    spectra with $S/N>10$ and $F_{H}>0.1$ (solid histogram).  The dashed
    histogram shows the $\phi_{H}$ distribution for normal galaxies,
    found by \citet{yip04a}.  \label{F_thetaHist}}
\end{figure}

%
\begin{figure*}[t]
  \plotone{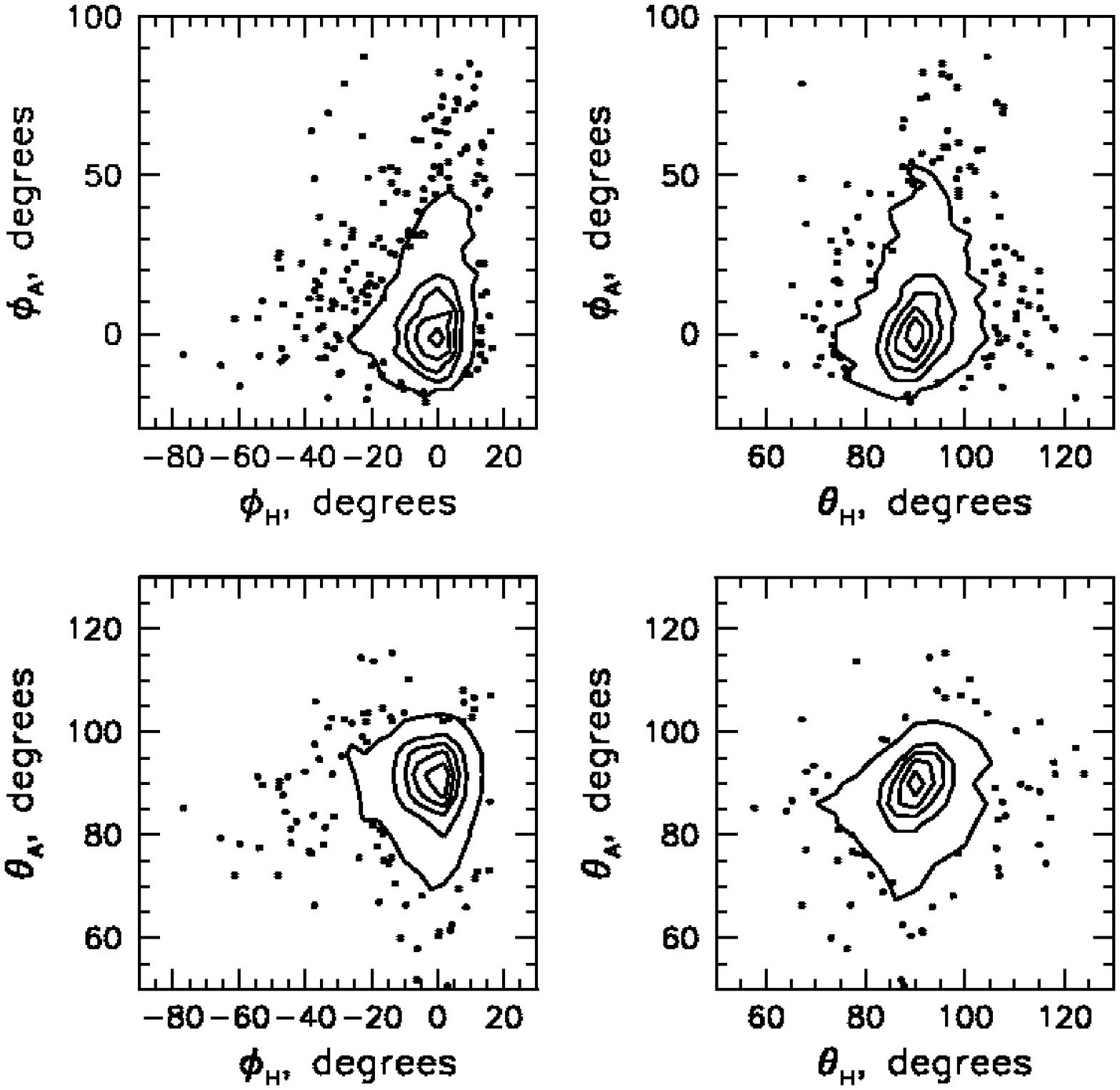}
  \caption{The relationships among the AGN and host galaxy classification
    angles.  Contours are spaced by density differences of approximately
    $20\%$ of the peak value.  Points are shown if they fall, or nearly
    fall, outside of the lowest density contour.  There are mild but
    significant correlations among the classification angles, except
    for $\theta_{A}$ vs.\ $\phi_{H}$.  \label{F_phiThetaAH}}
\end{figure*}

There is a much more dramatic difference in the distributions
of the second galaxy classification angle, $\theta_{H}$, shown
in Fig.\,\ref{F_thetaHist}.  The inactive galaxy distribution is
strongly peaked at $\theta_{H}\approx 89\arcdeg$, while the host
galaxy distribution is not only shifted to higher angles (a peak at
$\theta_{H}\approx 93\arcdeg$), but is much wider, and extends to
angles at which there are virtually no inactive galaxies.  A possible
interpretation is that the population of AGN hosts contains a relatively
large fraction of post-starburst galaxies.  Post-starburst (also called
E+A) galaxies are characterized by Balmer jumps and high-order Balmer
absorption lines, indicative of stellar populations of order 100 Myr old,
but with little or no narrow emission lines that would imply current
star formation \citep{dressler83}.  The third galaxy eigenspectrum has
properties similar to those of post-starburst galaxies \citep{yip04a}:
a blue continuum and strong Balmer absorption features, implying a
relatively young stellar population, but also features anticorrelated
with the continuum which reduce the narrow emission lines, and therefore
indicate suppressed star formation.  Post-starburst activity may therefore
be indicated by $\theta_{H}$ which is a measure of the relative strength
of the third galaxy eigenspectrum.  A strong third eigenspectrum does
not mean that there is no current star formation in a galaxy, nor that
the host galaxy can be classified as a post-starburst, only that there is
a component of the galaxy spectrum that resembles a post-starburst galaxy.

Spectacular examples of post-starburst AGNs (broad-line AGNs
associated with a post-starburst host) have been studied in some detail
\citep{brotherton99, canalizo00, brotherton02}.  In a study of so-called
``transition'' AGNs --- objects with characteristics of both AGNs and
ultra-luminous infrared galaxies --- \citet{canalizo01} found nearly
all of the host galaxies had post-starburst spectra.  Post-starburst
AGNs virtually always show morphological evidence for recent mergers or
other interactions; however, the evolutionary stages of the AGNs and the
relation to the post-starburst activity is not clear.  If $\theta_{H}$
is an indicator of post-starburst activity which is somehow linked
to triggering AGNs, the link may be reflected in a relation between
$\theta_{H}$ and black hole mass, as the black holes in young AGNs
would be expected to be less massive than in older AGNs for a given
accretion rate.

Assuming the eigenspectra are physically meaningful, connections among
the AGN and host components may be evident in the eigencoefficients.
The AGN and host galaxy classification angles are plotted against each
other in Fig.\,\ref{F_phiThetaAH}.  The points have been restricted
to objects for which $F_{H} > 0.3$ and $S/N>10$, in order to include
only very reliable estimates of the angles.  There are significant
correlations among all of the classification angles, except for $\phi_{H}$
and $\theta_{A}$ with a Spearman rank correlation coefficient of $-0.019$
and a probability of about $32\%$ for 2678 degrees of freedom.
The three other classification angle pairs, $\phi_{H}$ vs.\ $\phi_{A}$,
$\theta_{H}$ vs.\ $\phi_{A}$, and $\theta_{H}$ vs.\ $\theta_{A}$, have rank
correlation coefficients $0.139$, $0.292$, and $0.321$, respectively,
and probabilities of occurring by chance of much less than $1\%$ in
each case.  The mild but significant correlations among the host galaxy
and quasar classification angles may link recent star formation to the
physical properties of the AGNs.

\section{Discussion \& Summary\label{S_discussion}}
The results presented here suggest that eigenspectrum decomposition
has the potential to be a powerful technique for investigating the
properties of broad-line AGNs and their host galaxies.  We have
shown that the technique can efficiently and reliably separate the
spectroscopic components in composite spectra under certain conditions.
The approach is similar in some respects to that of \citet{hao05a},
who used eigenspectra of pure absorption-line galaxies to remove the
stellar component of active galaxies.  Here we have used sets of both
galaxy and AGN eigenspectra to reconstruct each component, which allows
the analysis of all the continuum, emission, and absorption features.
A large number of broad-line AGN spectra in the SDSS dataset are suitable
for this technique, but there is no reason it cannot be used for any
set of optical/UV flux calibrated AGN spectra.  There are limitations
dependent on the $S/N$ ratio, the fractional contribution of the host
galaxy, and the redshift (\S\,3).  The redshift limit
is imposed only by the wavelength coverage of the galaxy eigenspectra,
so it should be possible to extend the technique to higher redshifts by
constructing eigenspectra with shorter wavelength coverage, or obtaining
AGN spectra at near-IR wavelengths.

The reconstructed spectra of over 4600 broad-line AGNs in the SDSS show
that the host galaxies span a wide range of spectral classes, but the
distributions of their classification angles show marked differences
from their inactive counterparts.  In particular, the values of the
second classification angle indicate strong evidence for post-starburst
activity in a large fraction of the hosts.  Assuming that the host
galaxies are predominantly ellipticals or otherwise bulge-dominated,
the host population appears to be drawn mainly from the high luminosity
end of the luminosity function.  The colors of the hosts as a function
of luminosity also become bluer than expected for inactive elliptical
or bulge-dominated galaxies, as the host luminosity increases.  These
results can be qualitatively explained by black-hole/bulge mass scaling
relationships evident at low redshift, along with a link between recent
star formation and the formation of AGNs.  For typical SDSS spectra,
the sensitivity limits of the technique allow us to probe up to AGN
Eddington ratios a few tenths.  At the highest $S/N$ levels, there are still
virtually no AGNs that appear to emit beyond the Eddington limit.

Our results show that AGNs can reside in host galaxies with a wide variety
of spectral classes, and generally confirm previous imaging studies
that show that host galaxies tend to have bluer colors than expected
for inactive galaxies.  The imaging and spectroscopic decomposition
techniques are complementary, and there is much more information in
the spectra than we have presented in this paper.  Our focus here has
mainly been to illustrate the eigenspectrum decomposition technique, and
to highlight some initial results of its application to SDSS broad-line
AGN data.  There are a number of directions for future work, including
estimates of black hole masses, the spectral properties of the galaxies
and AGNs, the luminosity functions of both the AGNs and host galaxies,
and variations of the object properties among classes.

\acknowledgments
Funding for the creation and distribution of the SDSS Archive has
been provided by the Alfred P. Sloan Foundation, the Participating
Institutions, the National Aeronautics and Space Administration,
the National Science Foundation, the U.S. Department of Energy, the
Japanese Monbukagakusho, and the Max Planck Society. The SDSS Web site
is http://www.sdss.org/.

The SDSS is managed by the Astrophysical Research Consortium (ARC)
for the Participating Institutions. The Participating Institutions are
The University of Chicago, Fermilab, the Institute for Advanced Study,
the Japan Participation Group, The Johns Hopkins University, the Korean
Scientist Group, Los Alamos National Laboratory, the Max-Planck-Institute
for Astronomy (MPIA), the Max-Planck-Institute for Astrophysics (MPA),
New Mexico State University, University of Pittsburgh, University of
Portsmouth, Princeton University, the United States Naval Observatory,
and the University of Washington. 

D.~E.~V.~B and D.~P.~S. are supported in part by NSF grant AST~03-07982.


\clearpage
\clearpage
\begin{deluxetable}{lcl}
\tablecolumns{3}
\tabletypesize{\small}
\tablecaption{AGN and Host Galaxy Information Table Format \label{tab1}}
\tablewidth{0pt}
\tablehead{
  \colhead{Column} &
  \colhead{Format} &
  \colhead{Description}
}
\startdata
 1 & A18   &  Object Designation  hhmmss.ss+ddmmss.s
              (J2000; truncated coordinates) \\
 2 & F11.6 &  R.A. (J2000, degrees) \\
 3 & F11.6 &  Dec. (J2000, degrees) \\
 4 & F7.4  &  Redshift \\
 5 & I5    &  Spectroscopic Plate Number \\
 6 & I4    &  Spectroscopic Fiber Number \\
 7 & I6    &  Modified Julian Date of Spectroscopic Observation \\
 8 & A12   &  Morphology (resolved or unresolved) \\
 9 & F6.2  &  Average Spectroscopic S/N Per Pixel in the i Band \\
10 & F7.3  &  g Point Spread Function (PSF) Magnitude \\
11 & F6.3  &  g PSF Magnitude Error \\
12 & F7.3  &  r Point Spread Function (PSF) Magnitude \\
13 & F6.3  &  r PSF Magnitude Error \\
14 & F7.3  &  i Point Spread Function (PSF) Magnitude \\
15 & F6.3  &  i PSF Magnitude Error \\
16 & F7.3  &  g Spectroscopic Magnitude \\
17 & F7.3  &  r Spectroscopic Magnitude \\
18 & F7.3  &  i Spectroscopic Magnitude \\
19 & F7.3  &  g Cmodel Magnitude \\
20 & F6.3  &  g Cmodel Magnitude Error \\
21 & F7.3  &  r Cmodel Magnitude \\
22 & F6.3  &  r Cmodel Magnitude Error \\
23 & F7.3  &  i Cmodel Magnitude \\
24 & F6.3  &  i Cmodel Magnitude Error \\
25 & F7.3  &  g Band Galactic Absorption (magnitudes) \\
26 & F7.3  &  r Band Galactic Absorption (magnitudes) \\
27 & F7.3  &  i Band Galactic Absorption (magnitudes) \\
28 & F7.2  &  g Band AGN Component Absolute Magnitude \\
29 & F7.2  &  r Band AGN Component Absolute Magnitude \\
30 & F7.2  &  g Band Host Galaxy Absolute Magnitude \\
31 & F7.2  &  r Band Host Galaxy Absolute Magnitude \\
32 & F7.2  &  g Band Correction Applied to AGN Absolute Magnitude \\
33 & F7.2  &  r Band Correction Applied to AGN Absolute Magnitude \\
34 & F7.2  &  g Band Correction Applied to Host Absolute Magnitude \\
35 & F7.2  &  r Band Correction Applied to Host Absolute Magnitude \\
36 & F5.2  &  Flux Fraction from 4160-4210A from Host Galaxy \\
37 & F6.1  &  AGN First Classification Angle (degrees) \\
38 & F6.1  &  AGN Second Classification Angle (degrees) \\
39 & F6.1  &  Host Galaxy First Classification Angle (degrees) \\
40 & F6.1  &  Host Galaxy Second Classification Angle (degrees) \\
\enddata
\end{deluxetable}

%
\clearpage
\topmargin 0.5in
\begin{landscape}
\begin{deluxetable}{rrrrrrrrrrrrrrrrrr}
\tablecolumns{18}
\tabletypesize{\scriptsize}
\tablecaption{AGN and Host Galaxy Information\tablenotemark{a} \label{tab2}}
\tablewidth{0pt}
\tablehead{
  \colhead{Object} &
  \colhead{$\alpha_{J2000}$} &
  \colhead{$\delta_{J2000}$} &
  \colhead{Redshift} &
  \colhead{Plate} &
  \colhead{Fiber} &
  \colhead{MJD} &
  \colhead{Morphology} &
  \colhead{$S/N_{i}$} &
  \colhead{$M_{g,A}$\tablenotemark{b}} &
  \colhead{$M_{r,A}$\tablenotemark{b}} &
  \colhead{$M_{g,H}$\tablenotemark{b}} &
  \colhead{$M_{r,H}$\tablenotemark{b}} &
  \colhead{$F_{H}$\tablenotemark{b}} &
  \colhead{$\phi_{A}$\tablenotemark{b}} &
  \colhead{$\theta_{A}$\tablenotemark{b}} &
  \colhead{$\phi_{H}$\tablenotemark{b}} &
  \colhead{$\theta_{H}$\tablenotemark{b}}\\
  \colhead{(SDSS J)} &
  \colhead{(deg)} &
  \colhead{(deg)} &
  \colhead{} &
  \colhead{} &
  \colhead{} &
  \colhead{} &
  \colhead{} &
  \colhead{} &
  \colhead{} &
  \colhead{} &
  \colhead{} &
  \colhead{} &
  \colhead{} &
  \colhead{(deg)} &
  \colhead{(deg)} &
  \colhead{(deg)} &
  \colhead{(deg)}
}
\startdata
000011.41$+145545.6$ &  0.047547 &  14.929353 & 0.4596 & 0750 & 499 & 52235 & unresolved &  9.88 & -22.39 & -22.73 & -22.55 & -22.97 & 0.17 &   4.7 &  93.9 &  -7.4 &  93.4 \\
000011.96$+000225.3$ &  0.049842 &   0.040372 & 0.4790 & 0387 & 200 & 51791 & unresolved & 18.35 &   0.00 &   0.00 &   0.00 &   0.00 & 0.00 &   0.0 &   0.0 &   0.0 &   0.0 \\
000043.95$-091134.9$ &  0.183138 &  -9.193035 & 0.4388 & 0650 & 538 & 52143 & unresolved & 17.77 & -23.38 & -23.73 & -22.77 & -23.25 & 0.08 &   2.2 &  92.3 &  -8.2 &  85.9 \\
000048.15$-095404.1$ &  0.200661 &  -9.901158 & 0.2057 & 0650 & 494 & 52143 &   resolved & 13.04 & -19.65 & -20.13 & -21.51 & -22.11 & 0.68 &  14.1 &  85.1 & -10.2 &  96.3 \\
000102.18$-102326.9$ &  0.259117 & -10.390822 & 0.2943 & 0650 & 166 & 52143 & unresolved & 18.56 & -22.20 & -22.77 & -22.33 & -21.57 & 0.19 &  -4.6 &  82.7 & -32.7 &  95.1 \\
\enddata
\tablenotetext{a}{Table\,\ref{tab2} is presented in its entirety in the
  electronic edition of the Astronomical Journal.  A portion is shown
  here regarding its form and content.  The full catalog contains 40 
  columns of information for 11648 broad-line AGNs.}
\tablenotetext{b}{Default values, set to zeros, are given for entries with a non-physical spectroscopic decomposition.}
\end{deluxetable}
\clearpage
\end{landscape}

\end{document}